\documentclass[twoside,12pt]{article}

\usepackage{epsfig}

\usepackage{graphicx}
\usepackage{amsmath}
\usepackage{amssymb}
\usepackage{array}
\usepackage{wrapfig}
\usepackage{units}

\newcommand{\be}{\begin{equation}}
\newcommand{\ee}{\end{equation}}
\newcommand{\bea}{\begin{eqnarray}}
\newcommand{\eea}{\end{eqnarray}}

\newcommand{\ba}{\begin{eqnarray}}
\newcommand{\ea}{\end{eqnarray}}

\newcommand\slurp[1]{#1}
\newcommand\sslurp[2]{#1{#2}}
{\catcode`/=\active \expandafter}%
\slurp{\newcommand/}{/\penalty1000\hskip0pt\relax}
{\catcode`:=\active \expandafter}%
\slurp{\newcommand:}{:\penalty1000\hskip0pt\relax}

\newcommand\addspace{\ifcat\nextchar a\spacefactor999. \else.\fi}
{\catcode`\.=\active \expandafter}%
\slurp{\newcommand.}{\futurelet\nextchar\addspace}

\usepackage[unicode]{hyperref} 
\ifx\href\undefined\def\href#1{}\fi
\ifx\texorpdfstring\undefined\def\texorpdfstring#1#2{#1}\fi

\newcommand\myslash{/} \newcommand\mycolon{:}
\newcommand\doi{{\catcode`/=\active \catcode`:=\active \expandafter}\sslurp\realdoi}
{\catcode`/=\active \catcode`:=\active \expandafter}%
\slurp{\newcommand\realdoi[1]{{\let/=\myslash \let:=\mycolon
                               \edef\raw{{http://dx.doi.org/#1}}\expandafter}%
                               \expandafter\href\raw{doi:#1}}}
\newcommand\eprint[2]{{\escapechar-1%
                       \edef\a{\expandafter\string\csname arXiv\endcsname}%
                       \edef\b{\expandafter\string\csname #1\endcsname}%
                       \edef\c{\expandafter\string\csname #2\endcsname}%
                       \edef\d{\noexpand\href{http://arXiv.org/abs/\c}}%
                       \ifx\a\b\expandafter\d\fi{\tt #1:#2}}}

\begin{document}

\title{\vspace{1cm} 
Lepton Flavor and Number Conservation, \\ and Physics Beyond the Standard Model}

\author{Andr\'{e} de Gouv\^{e}a$^1$  and Petr Vogel$^2$
\\
\\
$^1$ Department of Physics and Astronomy, Northwestern University,\\
Evanston, Illinois, 60208, USA 
\\
$^2$ Kellogg Radiation Laboratory, Caltech, \\
Pasadena, California, 91125, USA
}

\maketitle

\begin{abstract}
The physics responsible for neutrino masses and lepton mixing remains
unknown.
More experimental data are needed to constrain and guide possible
generalizations of the standard
model of particle physics, and reveal the mechanism behind nonzero
neutrino masses. Here, the physics associated with
searches for the violation of lepton-flavor conservation in charged-lepton
processes
and the violation of lepton-number conservation in nuclear physics
processes is summarized.
In the first part, several aspects of charged-lepton flavor violation are
discussed,
especially its sensitivity to new particles and interactions beyond the
standard model of particle physics.
The discussion concentrates mostly on rare processes involving muons and
electrons.
In the second part, the status of the conservation of total lepton number
is discussed. The discussion here concentrates on current and future
probes of this apparent law of Nature via searches for neutrinoless double
beta decay, which is also the most sensitive probe
of the potential Majorana nature of neutrinos.
\end{abstract}

\newpage

\section{Introduction}

In the absence of interactions that lead to nonzero neutrino masses,  the Standard Model Lagrangian 
is invariant under global $U(1)_e\times U(1)_{\mu}\times U(1)_{\tau}$ rotations of the lepton fields. 
In other words, if neutrinos are massless, individual lepton-flavor numbers -- electron-number, 
muon-number, and tau-number -- are expected to be conserved. Very concretely, individual 
lepton-flavor quantum numbers are assigned as follows: $\ell=e,\mu$, or $\tau$,  $\nu_\ell$ 
have $Q_{\ell}=+1$, $\bar{\ell}, \bar{\nu}_{\ell}$ have $Q_{\ell}=-1$, while all other fields have charge zero.
Naturally, the total lepton number associated to $U(1)_L$, the ``diagonal'' subgroup of $U(1)_e\times U(1)_{\mu}\times U(1)_{\tau}$
 is a strictly conserved quantity.

In the last decade a variety of neutrino oscillation experiments proved
beyond any doubt that neutrino flavors, i.e., individual lepton flavor numbers, are not conserved. 
After propagating a finite distance neutrino beams of a given
flavor (electron, muon, tau) contain no longer only neutrinos of the
initial flavor. That phenomenon has been indisputably established \cite{Beringer:1900zz} in
``disappearance'' experiments with atmospheric and accelerator neutrinos
($\nu_{\mu}\nrightarrow\nu_{\mu}$ and 
$\bar{\nu}_{\mu} \nrightarrow \bar{\nu}_{\mu}$),  solar neutrinos
($\nu_e  \nrightarrow \nu_{e}$) and
reactor neutrinos ($\bar{\nu}_e  \nrightarrow \bar{\nu}_{e}$) and in ``appearance'' experiments with
solar neutrinos ($\nu_e  \rightarrow \nu_{\mu,\tau}$). There is also more than three-sigma 
evidence for the appearance of a new flavor from studies of atmospheric (anti)neutrinos ($\nu_{\mu}\rightarrow\nu_{\tau}$) \cite{Abe:2012jj}.

The only consistent explanation of all of these phenomena is (i) (some of the) neutrino masses are nonzero and 
distinct, (ii) the weakly interacting flavor neutrinos $\nu_e, \nu_{\mu}$ and $\nu_{\tau}$
are nontrivial superpositions of the so called
mass eigenstate neutrinos $\nu_1, \nu_2$ and $\nu_3$ (i.e. neutrinos with well-defined
 masses $m_1$, $m_2$ and $m_3$, respectively). These superpositions
are described by a $3 \times 3 $ unitary mixing matrix.

Further corroborating this picture, very recently, reactor experiments at distances corresponding to the
``atmospheric oscillation length'' \cite{DChooz,DayBay,RENO} as well as
the accelerator ``appearance'' ($\nu_{\mu} \rightarrow \nu_e$) experiments
\cite{MINOS,T2K} have successfully determined the magnitude of the until now
missing mixing angle $\theta_{13}$. The values of all three mixing
angles $\theta_{12} \sim \theta_{solar}$, $\theta_{23} \sim \theta_{atm}$ and
$\theta_{13}$ as well as  the two independent mass-squared differences 
$\Delta m_{21}^2 \sim \Delta m^2_{sol}$ and  
$\Delta m^2_{atm} \sim |\Delta m^2_{31}| \sim |\Delta m^2_{32}|$ are all
known now with good accuracy (see, for example, \cite{Beringer:1900zz,Tortola:2012te,Fogli:2012ua} for very recent summaries). 
Potential CP-violating parameters in the leptonic mixing matrix, along with the so-called neutrino mass hierarchy, remain virtually unconstrained.

\begin{figure}[tb]
\begin{center}
\begin{minipage}[t]{8cm}
\centerline{\epsfig{file=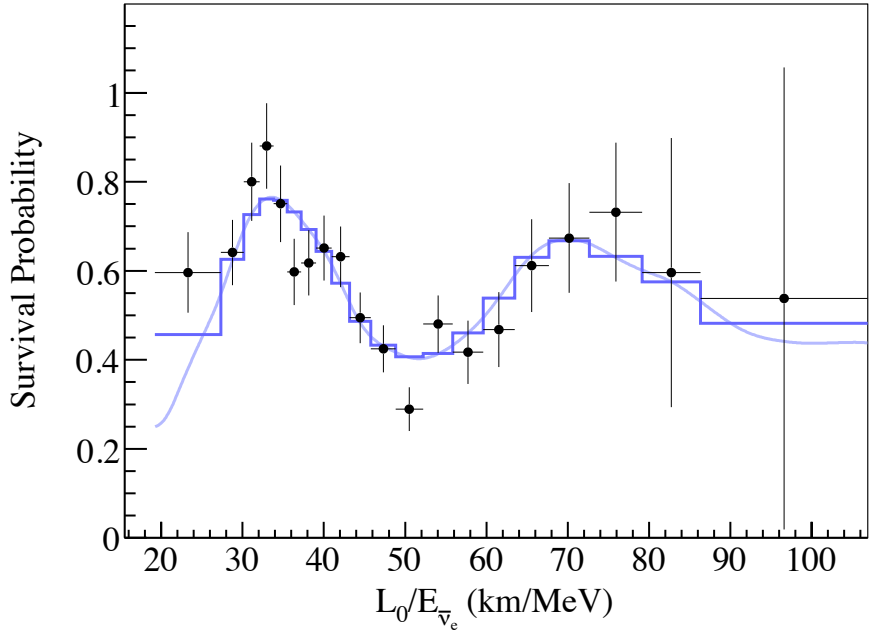,scale=0.5}}
\end{minipage}
\begin{minipage}[t]{16.5cm}
\caption{Ratio of the measured $\bar{\nu}_e$ flux to the no-oscillation 
expectation as the function of $L_0/E_{\nu}$. $L_0$ = 180 km is the 
flux weighted average distance from all nuclear reactors. The data points are background  and
geoneutrino subtracted with statistical errors only. The curve  shows the expectation from neutrino oscillations,
taking into account the different distances to individual reactors, time variations of the fluxes, and efficiencies.}
\label{fig:LE}
\end{minipage}
\end{center}
\end{figure}

The success of the neutrino oscillation hypothesis is illustrated in Fig. \ref{fig:LE},
based on the results of the KamLAND reactor oscillation experiment \cite{KamL}. 
This textbook-type example
clearly shows that the initial beam of electron antineutrinos changes into a flavor that is invisible to the detector, i.e., 
oscillates periodically as a function of $L/E_{\nu}$, where $L$ is the distance from the neutrino source 
and $E_{\nu}$ is the neutrino energy. 


The discovery of neutrino oscillations, and its implication of nonvanishing neutrino
masses and the non-conservation of neutrino flavor, implies that the Standard Model
is incomplete and needs generalization. The new physics responsible for neutrino masses and mixing remains unknown and
will only be revealed once more experimental data, from different areas of fundamental physics research,
become available. Here, we summarize the physics associated with 
searches for the violation of lepton-flavor conservation in charged-lepton processes, 
and the violation of lepton-number conservation in nuclear physics processes. While these two issues seemingly involve different aspects 
of the physics beyond the Standard Model, they are both intimately related to the understanding of the origin of neutrino masses,
and hence are discussed together.

The established existence of neutrino flavor violation implies that, barring accidental 
cancellations, all so-called lepton flavor violating processes are also allowed and will 
occur at some order in perturbation theory. The rates for such processes, however, cannot be estimated model-independently and
are hence expected to provide non-trivial information regarding the nature of new physics. 

On the other hand, neutrino oscillations do not necessarily imply that total lepton number is violated. 
While nonzero neutrino masses and lepton mixing imply that $U(1)_e\times U(1)_{\mu}\times U(1)_{\tau}$ is not a good global symmetry
of nature, $U(1)_L$ may still remain as an exact global symmetry of the ``New Standard Model'' Lagrangian (one that includes nonzero neutrino masses). This is the case if neutrinos turn out to be Dirac fermions. The situation here is analogous to that of the quark sector, where individual quark-flavor numbers are known to be violated by weak processes while baryon number $U(1)_B$ remains a good quantum number (at least as far as the standard model and all current observations are concerned).\footnote{Both $U(1)_B$ and (potentially) $U(1)_L$ are only ``classical'' symmetries of the new Standard Model Lagrangian, being violated at the quantum level. $U(1)_{B-L}$, however, is a potential nonanomalous global symmetry of nature even when nonzero neutrino masses are taken into account.} If neutrinos are Majorana fermions, however, $U(1)_L$ is not a good symmetry of the New Standard Model. Current data are silent concerning the nature -- Majorana fermions or Dirac fermions -- of the massive neutrinos. Different new physics models for nonzero neutrino masses predict the neutrinos to be Dirac or Majorana fermions. Only by investigating the validity of lepton-number conservation will we be able to distinguish one ``type'' of new physics from the other.

The paper is divided into two sections. Section~\ref{CLFV} deals with searches for charged-lepton flavor
violation. Charged-lepton flavor violation (CLFV) is 
defined as all charged-lepton processes that violate lepton-flavor number. These 
include $\ell\to\ell'\gamma$, $\ell\to\ell'\ell''\bar{\ell}'''$, $\ell+X\to\ell'+X$ and 
$X\to \ell\bar{\ell}'$, where $\ell,\ell',\ldots\in\{e,\mu,\tau\}$ and $X$ are states that 
carry no lepton-flavor number. We will discuss several aspects of CLFV, 
especially their sensitivity to new particles and interactions beyond the Standard Model. 
We will concentrate on rare processes involving muons and electrons, and will briefly 
comment on processes involving tau-leptons. There are several more comprehensive 
recent overviews of different aspects of CLFV, 
including \cite{reviews,Aysto:2001zs,deGouvea:2010zz,Kuno:1999jp}.
Section~\ref{0nubb} deals with the total lepton conservation (LNV), and in particular with the
probe of that law using neutrinoless double beta decay processes. Searches for the violation of lepton number 
often involve nuclear probes and the analyses of the corresponding
experiments often require deep understanding of nuclear structure. We will briefly
review such problems, but the main emphasis will be on the possible new physics
origins of the violation of this conservation law.

\section{Lepton Flavor Conservation}
\label{CLFV}

\subsection{Muon to Electron Searches: Current Bounds, Near Future Goals}


Searches for $\mu\to e\gamma$ date back to the late 1940's. By the early 1960's it was clear that naive estimates for the branching ratio for $\mu\to e\gamma$ from the physics responsible for $\mu\to e\nu\bar{\nu}$ were violated by negative experimental results. Indeed, the fact that $\mu\to e\gamma$ did not occur with branching ratios above  $10^{-6}$ was among the most pressing arguments in favor of the so-called `two-neutrino hypothesis' and the postulate that there were two lepton flavors, each accompanied by a conserved lepton-flavor number.\footnote{The hypothesis was expanded in the 1970's to include a third conserved lepton-flavor number.} In modern textbooks, we describe (negative) muon decay as $\mu^-\to e^-\bar{\nu}_e\nu_{\mu}$; the electron neutrino and the muon neutrino are different objects.

Along with $\mu^+\to e^+\gamma$ decays, two other rare muon processes provide the most stringent current bounds on CLFV, and promise the most near-future sensitivity to these phenomena: $\mu^+ \to e^+e^-e^+$ decays, and $\mu^-\to e^-$~conversion in nuclei. The three are briefly discussed here. Other CLFV processes involving muons and electrons include rare kaon decays like $K_L\to\mu^{\pm}e^{\mp}$, $K\to\pi\mu^{\pm}e^{\mp}$, and muonimum--antimuonium oscillations, $\mu^+e^-\leftrightarrow\mu^-e^+$  \cite{Beringer:1900zz}. Some of these are very severely constrained and sensitive to different types of new physics at different levels. Nonetheless, they will not be considered henceforth.

$\mu^+\to e^+\gamma$ decays are currently being pursued by the MEG experiment at the Paul Scherrer Institut (\cite{Adam:2011ch} and references therein). Presently,  Br$(\mu^+\to e^+\gamma)<2.4\times 10^{-12}$ at the 90\% confidence level \cite{Adam:2011ch}. MEG aims at being sensitive to Br$(\mu^+\to e^+\gamma)\gtrsim 10^{-13}$, while potential improvements may lead to sensitivities approaching $10^{-14}$ \cite{future_meg}.\footnote{After the completion of this review, the MEG experiments made public their most recent result, Br$(\mu^+\to e^+\gamma)<5.7\times 10^{-13}$ at the 90\% confidence level \cite{Adam:2013mnn}.} Experimentally, it is extremely challenging to envision improving the sensitivity to Br$(\mu\to e\gamma)$ beyond $10^{-14}$ (for a discussion, see \cite{Aysto:2001zs}) -- the current sensitivity is already dominated by accidental backgrounds which grow with the muon beam intensity. For this reason, there are no proposals for improving on the sensitivy to $\mu\to e\gamma$ beyond what can be ultimately achieved by MEG. 

Presently, the Br$(\mu^+\to e^+e^-e^+)<1.0\times 10^{-12}$ at the 90\% confidence level \cite{Bellgardt:1987du}. This limit was obtained over 20 years ago. Presently, a letter of intent has been submitted to the Paul Scherrer Institut to conduct an improved search for $\mu^+\to e^+e^-e^+$, ultimately sensitive to branching ratios above $10^{-15}$ or, perhaps, $10^{-16}$ \cite{mu3e}. As with $\mu^+\to e^+\gamma$ decays, sensitivity beyond $10^{-16}$, regardless of the intensity of the muon beam, appears very challenging.

$\mu^-\to e^-$~conversion in nuclei is the process where a bound muon interacts with the nucleus and converts into an electron, which is born with enough kinetic energy to ``escape'' the Coulomb potential: $\mu^- N\to e^-N^{(*)}$, where $N$ is some nucleus. Experimentally, the signal for $\mu^-\to e^-$~conversion in nuclei is a monochromatic electron, whose energy lies just beyond the kinematical end-point of Michel electrons produced by muon decay in orbit. The $\mu^-\to e^-$~conversion rate is usually expressed in units of the capture rate, $\mu^- N\to \nu_{\mu}N'$. Presently, the most stringent bound is Br$(\mu\to e~{\rm conv~in~Au})<7\times 10^{-13}$ at the 90\% confidence level \cite{Bertl:2006up}. There are  a few proposals aimed at experiments sensitive to Br$(\mu\to e~{\rm conv~in~Al})> 10^{-16}$ or better\footnote{Some proposals involve other nuclei, including $^{48}$Ti.} by the end of this decade \cite{Carey:2008zz,Kurup:2011zza}. Unlike the two rare decays discussed earlier, searches for $\mu^-\to e^-$~conversion in nuclei are not, naively, expected to hit any experimental ``wall'' until conversion rates below $10^{-18}$ or lower  \cite{Aysto:2001zs}. Experimental setups sensitive to conversion rates below $10^{-17}$ are currently under serious study both at Fermilab (assuming Project X becomes available) and J-PARC. Experimentally, in the long-run, it is widely anticipated that $\mu^-\to e^-$~conversion in nuclei will provide the ultimate sensitivity to CLFV.

\subsection{$\nu$ Standard Model Expectations}
In spite of the fact that we have determined that CLFV must occur, measurements of neutrino oscillation processes do not allow us to reliably estimate the rate for the various CLFV processes. The reason is that while neutrino oscillation phenomena depend only on neutrino masses and lepton mixing angles, the rates for the various CLFV processes depend dramatically on the mechanism behind neutrino masses and lepton mixing, currently unknown (for recent reviews see, for example, \cite{Mohapatra:2005wg}). Different neutrino mass-generating Lagrangians lead to very different rates for CLFV. Some of these will be discussed briefly here and in Sec.~\ref{sec:example}.

The massive neutrino contribution to CLFV that involves only active neutrinos is absurdly small. For example \cite{Marciano:1977wx},
\begin{equation}
{\rm Br}(\mu \to e \gamma )=\frac{3\alpha}{32 \pi}\left|\sum_{i=2,3} 
U_{\mu i}^*U_{ei} \frac{\Delta m_{i1}^2}{M_W^2}\right|^2 < 10^{-54}\,,
\label{meg_sm}
\end{equation}
where $U_{\alpha i}$ are elements of the neutrino mixing matrix, $\Delta m^2_{ij}$ are the neutrino mass-squared differences, $\alpha$ is the fine-structure constant, and $M_W$ is the $W$-boson mass. Similar ridiculously small rates are expected for $\mu\to eee$, $\mu\to e$~conversion and rare process involving taus. The estimate above applies to some neutrino mass models, including minimal scenarios with Dirac neutrinos. The reason behind the tiny branching ratios is well-known. CLFV, as defined above, is a flavor-changing neutral current process and such processes are subject to the GIM mechanism (or generalizations thereof). 

In many neutrino-mass generating scenarios, the active neutrino contribution turns out to be, not surprisingly, severely subdominant. In the famous seesaw mechanism \cite{seesaw1,seesaw2}, for example, heavy neutrino contributions to CLFV are naively expected to be of order those of the light neutrinos, but there is no theorem that prevents them from being much, much larger. According to \cite{de Gouvea:2007uz}, for example, current experimental constraints on the seesaw Lagrangian allow Br($\tau\to\mu\gamma$) as large as $10^{-9}$,  Br($\mu\to e\gamma$) as large are $4\times 10^{-13}$, and normalized rates for $\mu\to e$ conversion in nuclei that saturate the current experimental upper bound. For more details, see, for examples, \cite{Dinh:2012bp}.

\subsection{Some Model Independent Considerations}

Independent from the mechanism behind neutrino masses, it is often speculated that the rates for different CLFV processes are, perhaps, just beneath current experimental upper bounds. The reason is we suspect, for several reasons, that there are new degrees of freedom beyond those in the Standard Model. We also suspect that some of those have masses around 1~TeV. Since lepton-flavor numbers are known not to be good quantum numbers, it is generically expected that virtual processes involving the new degrees of freedom will mediate, at some order in perturbation theory, CLFV. 

Some concrete new physics scenarios will be briefly discussed in Sec.~\ref{sec:example}. Here we discuss two effective Lagrangians\footnote{Parts of this discussion were first presented in writing in \cite{Carey:2008zz,deGouvea:2010zz}.} that mediate CLFV processes involving muons aiming at illustrating how searches for CLFV are sensitive to new physics, and how different CLFV channels compare with one another. 

After integrating out heavy degrees of freedom, and after electroweak symmetry breaking, CLFV is mediated by effective operators of dimension five and higher. We first concentrate on the following effective Lagrangian\footnote{The most general effective Lagragian includes several other terms \cite{Kuno:1999jp}. The subsets included in Eqs.~(\ref{l_mec},\ref{eq:l_meee}), however, are sufficient to illustrate all issues discussed here. Modulo extreme constructive/destructive interference effects among different effective operators, the points made here remain valid.}
\begin{eqnarray}
{\cal L}_{\rm CLFV}=\frac{m_{\mu}}{(\kappa+1)\Lambda^2}\bar{\mu}_R\sigma_{\mu\nu}e_LF^{\mu\nu}+ h.c. \nonumber \\ 
\frac{\kappa}{(1+\kappa)\Lambda^2}\bar{\mu}_L\gamma_{\mu}e_L\left(\bar{u}_L\gamma^{\mu}u_L+\bar{d}_L\gamma^{\mu}d_L\right)+ h.c.\,.
\label{l_mec}
\end{eqnarray}
The subscripts $L,R$ indicate the chirality of the different Standard Model fermion fields, $F^{\mu\nu}$ is the photon field strength and $m_{\mu}$ is the muon mass. The coefficients of the two types of operators are parameterized by two independent constants: the dimensionful $\Lambda$ parameter (with dimensions of mass), which is meant to represent the effective mass scale of the new degrees of freedom, and the dimensionless parameter $\kappa$, which governs the relative size of the two different types of operators. The magnetic-moment type operator in the first line of Eq.~(\ref{l_mec}) directly mediates $\mu\to e\gamma$ and mediates $\mu\to eee$ and $\mu\to e$~conversion in nuclei at order $\alpha$. The four-fermion operators in the second line of Eq.~(\ref{l_mec}), on the other hand, mediate $\mu\to e$~conversion at the leading order and $\mu\to e\gamma$, $\mu\to eee$ at the one-loop level. For $\kappa\ll 1$, the dipole-type operator dominates CLFV phenomena, while for $\kappa\gg 1$ the four-fermion operators are dominant.

\begin{figure}[tb]
\begin{center}
\begin{minipage}[t]{8 cm}
\centerline{\epsfig{file=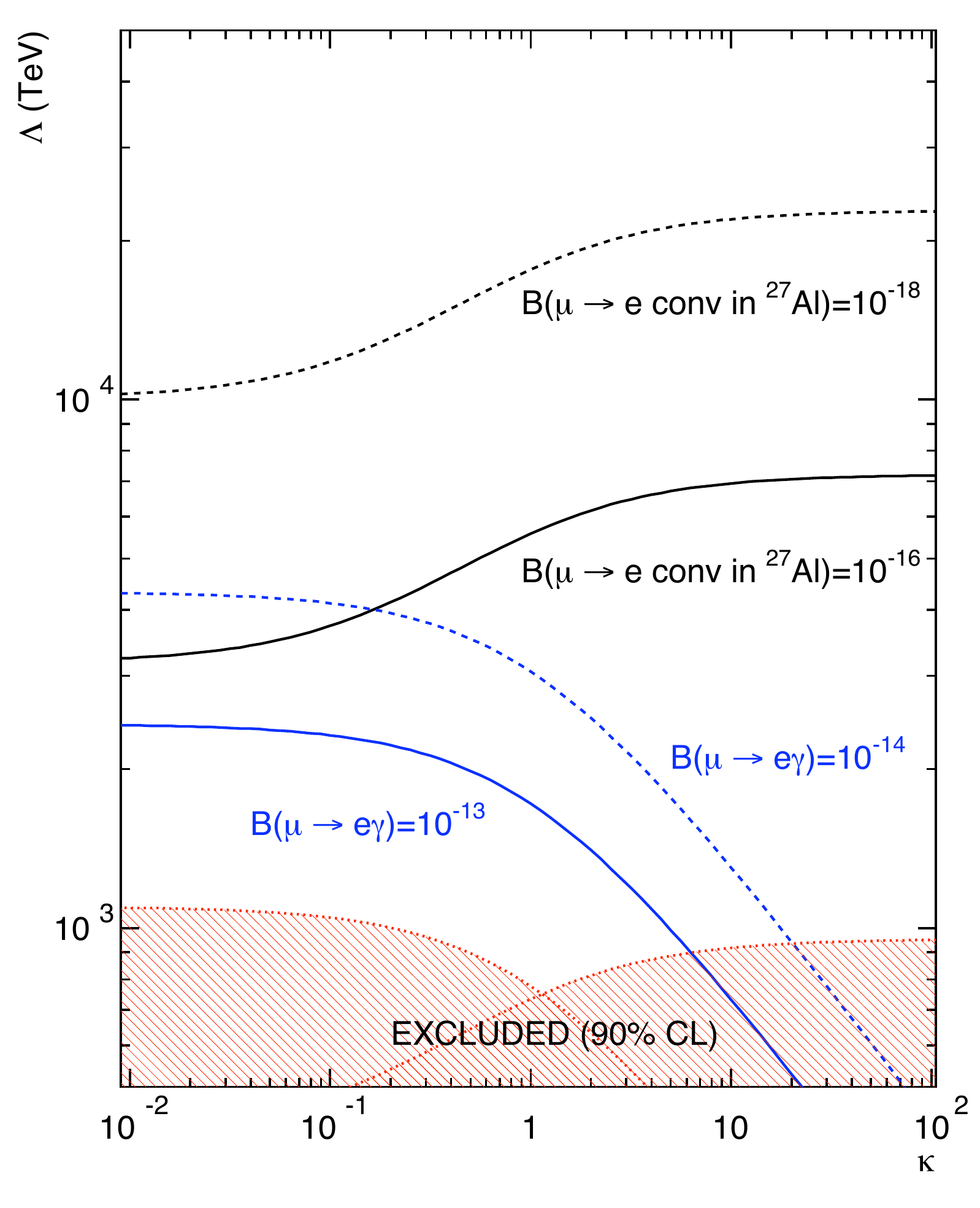,scale=0.7}}
\end{minipage}
\begin{minipage}[t]{16.5 cm}
\caption{Sensitivity of a $\mu\to e$~conversion in $^{27}$Al experiment that can probe a normalized capture rate of $10^{-16}$ and $10^{-18}$, and of a $\mu\to e\gamma$ search that is sensitive to a branching ratio of $10^{-13}$ and $10^{-14}$, to the new physics scale $\Lambda$ as a function of $\kappa$, as defined in Eq.~(\ref{l_mec}). Also depicted is the currently excluded region of this parameter space.}
\label{fig:meg_mec}
\end{minipage}
\end{center}
\end{figure}

The sensitivity to $\Lambda$ as a function of $\kappa$ for $\mu\to e\gamma$ and $\mu\to e$~conversion efforts is depicted in Fig.~\ref{fig:meg_mec}. For $\kappa\ll1$, an experiment sensitive to Br$(\mu\to e\gamma)>10^{-13}$ will probe $\Lambda$ values less than 2500~TeV, while for $\kappa\gg1$ an experiment sensitive to Br$(\mu\to e~{\rm conv~in~^{27}Al})>10^{-16}$ will probe $\Lambda$ values less than 7000~TeV.

Relevant information can be extracted from Fig.~\ref{fig:meg_mec}. CLFV already probes $\Lambda$ values close to 1000~TeV and next-generation experiments will start to probe $\Lambda\sim 10^4$~TeV and beyond. Furthermore, a $\mu\to e$~conversion experiment is ``guaranteed'' to outperform a $\mu\to e\gamma$ experiment for any value of $\kappa$ as long as it is a couple of orders of magnitude more sensitive. Since, as already discussed, it appears very challenging to perform a $\mu\to e\gamma$ experiment sensitive to branching ratios smaller than $10^{-14}$, $\mu\to e$~conversion searches (not expected to hit any ``wall'' before normalized rates around at most $10^{-18}$), are the more effective way of pursuing CLFV after the on-going MEG experiment is done analyzing its data.

Similarly, we can ask what are the consequences for CLFV if the new physics is best captured by the following ``leptons-only'' effective Lagrangian:
\begin{eqnarray}
{\cal L}_{\rm CLFV}= \frac{m_{\mu}}{(\kappa+1)\Lambda^2}\bar{\mu}_R\sigma_{\mu\nu}e_LF^{\mu\nu}+ h.c. \nonumber \\
\frac{\kappa}{(1+\kappa)\Lambda^2}\bar{\mu}_L\gamma_{\mu}e_L\left(\bar{e}\gamma^{\mu}e\right)+ h.c. \,.
\label{eq:l_meee}
\end{eqnarray}
Similar to the dimension-six operators in the second line of Eq.~(\ref{l_mec}), the dimension-six operator in the second line of Eq.~(\ref{eq:l_meee}) mediates $\mu\to eee$ at the tree level and $\mu\to e\gamma$, $\mu\to e$~conversion at the one-loop level. Similar to Eq.~(\ref{l_mec}), the dimensionless parameter $\kappa$ determines whether the dipole-like or the four-fermion interaction is dominant when it comes to CLFV.

\begin{figure}[tb]
\begin{center}
\begin{minipage}[t]{8 cm}
\centerline{\epsfig{file=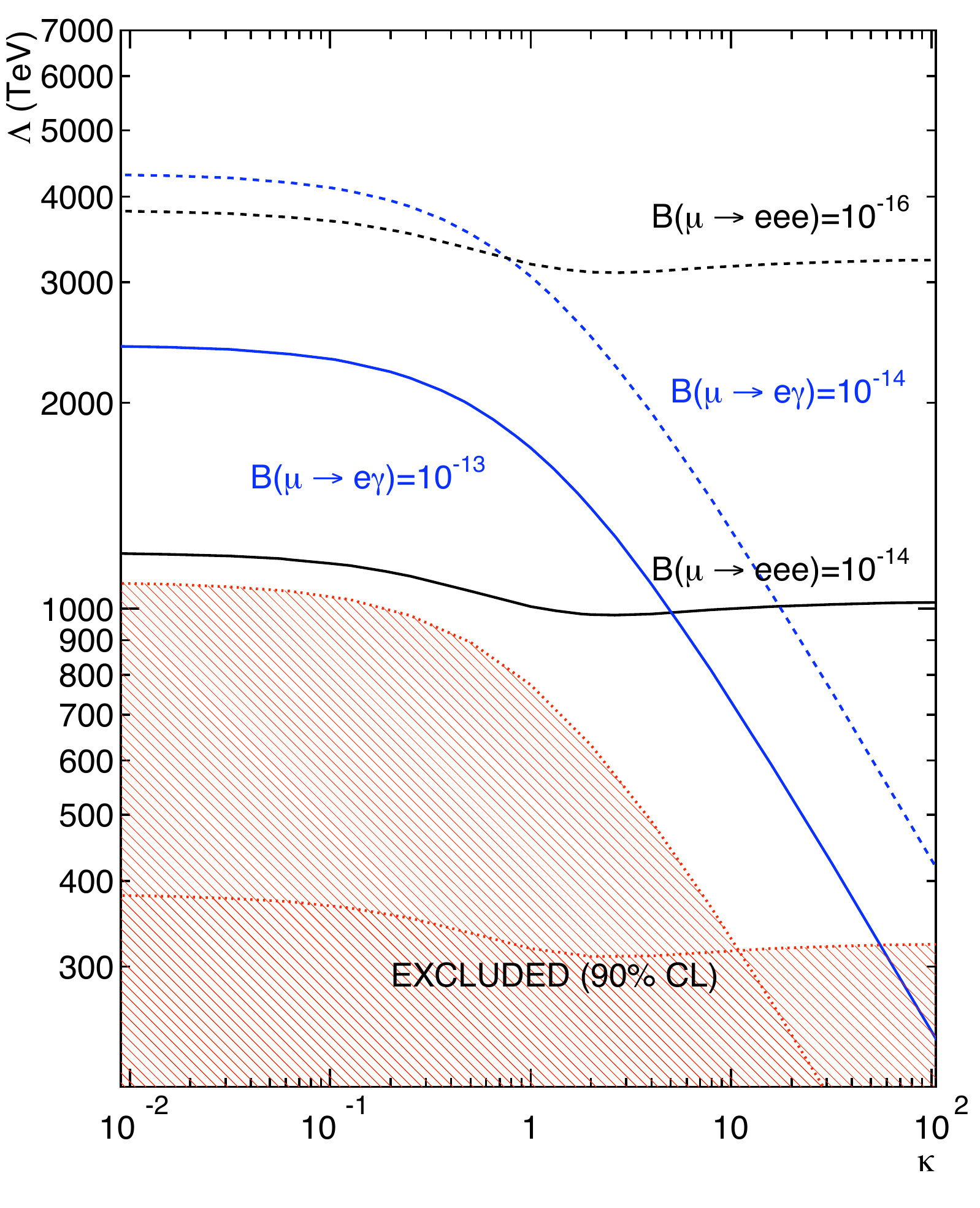,scale=0.7}}
\end{minipage}
\begin{minipage}[t]{16.5 cm}
\caption{Sensitivity of a $\mu\to eee$ experiment that is sensitive to branching ratios $10^{-14}$ and $10^{-16}$, and of a $\mu\to e\gamma$ search that is sensitive to a branching ratio of $10^{-13}$ and $10^{-14}$, to the new physics scale $\Lambda$ as a function of $\kappa$, as defined in Eq.~(\ref{eq:l_meee}). Also depicted is the currently excluded region of this parameter space.}
\label{fig:meg_meee}
\end{minipage}
\end{center}
\end{figure}

The sensitivity to $\Lambda$ as a function of $\kappa$ for $\mu\to e\gamma$ and $\mu\to eee$ efforts is depicted in Fig.~\ref{fig:meg_meee}. Here, for $\kappa\gg1$, an experiment sensitive to Br$(\mu\to eee)>10^{-15}$ will probe $\Lambda$ values less than 1800~TeV. As in the example depicted in Fig.~\ref{fig:meg_mec}, we note that a $\mu\to eee$ experiment is guaranteed to outperform a $\mu\to e\gamma$ experiment, for any value of $\kappa$, as long as it is a few hundred times more sensitive. Whether this can be realistically achieved in future experiments is still under investigation \cite{Aysto:2001zs,mu3e}.

A model independent comparison between the reach of $\mu\to eee$ and $\mu\to e$~conversion in nuclei is a lot less straight forward. If the new physics is such that the dipole-type operator is dominant ($\kappa\ll 1$ in Figures~\ref{fig:meg_mec} and \ref{fig:meg_meee}), it is easy to see that near-future prospects for $\mu\to e$~conversion searches are comparable to those for $\mu\to eee$, assuming both can reach the $10^{-16}$ level. $\mu\to e$~conversion searches will ultimately dominate, assuming these can reach beyond $10^{-17}$, and assuming $\mu\to eee$ searches ``saturate'' at the $10^{-16}$ level. Under all other theoretical circumstances, keeping in mind that $\kappa$ and $\Lambda$ in  Eqs.~(\ref{l_mec},\ref{eq:l_meee}) are {\em not} the same, it is impossible to unambiguously compare the two CLFV probes. 

The discussions above also serve to illustrate another ``feature'' of searches for CLFV violation. In the case of a positive signal, the amount of information regarding the new physics is limited. For example, a positive signal in a $\mu\to e$~conversion experiment does not allow one to measure either $\Lambda$ or $\kappa$ but only a function of the two. In order to learn more about the new physics, one needs to combine information involving the rate of a particular CLFV process with other observables. These include other CLFV observables ({\it e.g.}, a positive signal in $\mu\to e\gamma$ and $\mu\to eee$ would allow one to measure both $\kappa$ and $\Lambda$ if Eq.~(\ref{eq:l_meee}) describes CLFV), studies of electromagnetic properties of charged leptons ($g-2$, electric dipole moments), precision studies of neutrino processes (including oscillations), and, of course, ``direct'' searches for new, heavy degrees of freedom (Tevatron, LHC). Valuable information, including the nature and chirality of the effective operators that mediate CLFV, can be obtained by observing $\mu\to e$~conversion in different nuclei \cite{Kuno:1999jp,Kitano:2002mt,Cirigliano:2009bz} or by studying the kinematical distribution of the final-state electrons in $\mu\to eee$ (see \cite{Kuno:1999jp} and references therein).

Before moving on to specific new physics scenarios, it is illustrative to compare, as model-independently as possible, new physics that mediates CLFV and the new physics that may have manifested itself in precision measurements of the muon anomalous magnetic moment. In a nutshell, the world's most precise measurement of the $g-2$ of the muon disagrees with the world's best Standard Model estimate for this observable at the $3.6\sigma$ level (for an updated overview see \cite{Beringer:1900zz}, and references therein). New, heavy physics contributions to the muon $g-2$ are captured by the following effective Lagrangian:
\begin{equation}
{\cal L}_{g-2}\supset \frac{m_{\mu}}{\Lambda^2}\bar{\mu}_R\sigma_{\mu\nu}\mu_L F^{\mu\nu}+h.c. \,.
\label{l_g-2}
\end{equation}
Current $g-2$ data point to $\Lambda\sim 8$~TeV.  Eq.~(\ref{l_g-2}), however, is very similar to Eqs.~(\ref{l_mec},\ref{eq:l_meee}) in the limit $\kappa\ll 1$, keeping in mind that $\Lambda$ in Eq.~(\ref{l_g-2}) need not represent the same quantity as $\Lambda$ in Eqs.~(\ref{l_mec},\ref{eq:l_meee}) in the limit $\kappa\ll 1$.

We can relate $\Lambda$ in Eq.~(\ref{l_g-2}) to that in Eqs.~(\ref{l_mec},\ref{eq:l_meee}) in the following suggestive way: $(\Lambda_{\rm CLFV})^{-2}=\theta_{e\mu}(\Lambda_{g-2})^{-2}$. Here the parameter $\theta_{e\mu}$ measures how flavor-conserving is the new physics. For example, if $\theta_{e\mu}=0$, the new physics is strictly flavor conserving, while if the new physics is flavor-indiferent, $\theta_{e\mu}\sim 1$. In the latter case, negative searches for $\mu\to e\gamma$ already preclude a new physics interpretation to the muon $g-2$ results as these constrain $\Lambda\gtrsim1000$~TeV. On the other hand, if the muon $g-2$ discrepancy is really evidence for new physics, searches for $\mu\to e\gamma$ reveal that the ``amount'' of flavor violation in the new physics sector is very small: $\theta_{e\mu}<10^{-4}$. A more detailed comparison of these two probes of new physics can be found, for example, in \cite{Hisano:2001qz}.

\subsection{Some Scenarios and Examples}
\label{sec:example}

Models for new physics that introduce new degrees of freedom at the weak scale are often severely constrained by, or make specific predictions for, CLFV observables. These include new physics scenarios that address the gauge hierarchy problem and certain models for nonzero neutrino masses. Concrete predictions further require that the models specify the flavor structure of the new physics, and are also obstructed by the fact that some of the flavor parameters of the lepton sector, including the neutrino mass hierarchy and the value of all CP-violating parameters, remain unknown. Here we discuss, in a little detail, a couple of examples aimed at conveying the most important lessons and challenges, and list other possibilities. 

Neutrino Majorana masses can be generated if the Standard Model particle content is augmented to include a Higgs boson $SU(2)$ triplet $T$ with hypercharge one \cite{seesaw2}. In this case, the gauge symmetries allow 
\begin{equation}
{\cal L}\supset \lambda_{\alpha\beta}L_{\alpha}TL_{\beta}+h.c.,
\label{eq:ssII}
\end{equation}
where $\lambda_{\alpha\beta}=\lambda_{\beta\alpha}$ are dimensionless Yukawa couplings, $L_{\alpha}$ are the $SU(2)$ lepton doublet fields, $\alpha,\beta=e,\mu,\tau$. After the neutral component of $T$ acquires a vacuum expectation values $v_T$, neutrinos acquire a Majorana mass matrix $m_{\nu}=\lambda v_T$. 

All CLFV processes are mediated by one-loop diagrams involving the Higgs triplet fields, while $\ell\to\ell'\ell''\ell'''$ processes are mediated at the tree level via the exchange of the doubly-charged components of $T$ (for details see, for example, \cite{Chun:2003ej,Dinh:2012bp} and references therein). This scenario manifests itself at low-energies in the form of (a variant of) Eq.~(\ref{eq:l_meee}) with, in general, large $\kappa$. 

The branching ratio for $\mu\to eee$, for example, is
\begin{equation}
{\rm Br}(\mu\to eee)\propto\frac{M_W^2}{M^2_T}\frac{|(m_{\nu}^*)_{\mu e}(m_{\nu})_{ee}|^2}{v_T^4},
\end{equation} 
where $M_T$ is the mass of the doubly-charged component of $T$. The rates for other CLFV observables are also functions of $v_T$, $M_T$, and the elements of $m_{\nu}$. Concrete predictions for the rates of CLFV processes, in this case, are not possible without more external assumptions;  there is no input regarding $M_T$ and $v_T$ (unless triplet Higgs bosons are discovered at the LHC!). Correlations among different CLFV observables, however, are well-defined; $m_{\nu}$ is, in principle, uniquely determined by neutrino masses and lepton mixing observables. Recent detailed studies, which take into account the recent discovery that $\theta_{13}$ is not zero, can be found in \cite{Dinh:2012bp,Chakrabortty:2012vp}.

A class of very well studied models that usually predicts $\kappa\ll 1$ is that of supersymmetric versions of the Standard Model with R-parity conservation. Superpartner loops involving gauginos and sleptons lead to \cite{Borzumati:1986qx}, in the language of Eqs.~(\ref{l_mec},\ref{eq:l_meee}),
\begin{equation}
\frac{1}{\Lambda^2}\sim \frac{g^2e}{16\pi^2M^2_{\rm SUSY}}\theta_{e\mu}\,,
\end{equation}
where $e,g$ are the electromagnetic and $SU(2)_L$ couplings, respectively, and $M_{\rm SUSY}$ is some effective superpartner mass. The flavor violating parameter $\theta_{e\mu}$ can be expressed in terms of the off-diagonal slepton mass-squared matrix element $\Delta m^2_{\tilde{\mu}\tilde{e}}$, 
\begin{equation}
\theta_{\mu e}\sim \frac{\Delta m^2_{\tilde{\mu}\tilde{e}}}{M^2_{\rm SUSY}}\,.
\end{equation}
Similar to the $g-2$ discussion in the previous section, weak-scale superpartner masses are only allowed by CLFV data if $\theta_{\mu e}$ is small, which, in this case, translates into  $\Delta m^2_{\tilde{\mu}\tilde{e}}$ much smaller than the diagonal slepton masses-squared. 

One way to guarantee that this is the case is to assume that the mechanism of supersymmetry breaking, which determines the slepton mass-squared matrix, is flavor blind so that, in the absence of Standard Model flavor effects, $\Delta m^2_{\tilde{\mu}\tilde{e}}\equiv 0$. Since, in the Standard Model with nonzero neutrino masses, flavor-numbers are not conserved, quantum mechanical effects lead to nonzero, calculable $\Delta m^2_{\tilde{\mu}\tilde{e}}$. Concrete predictions, alas, depend dramatically on the mechanism behind neutrino masses. In the case of the Type-I seesaw mechanism, for example,
\begin{equation}
\Delta m^2_{\tilde{\alpha}\tilde{\beta}}
\sim \frac{M^2_{\rm SUSY}}{8\pi^2} \sum_k
\left( y \right)^*_{\alpha k} \left( y \right)_{\beta k} \ln
\frac{M_{UV}}{M_{N_k}}\,,
\label{susy_seesaw}
\end{equation}
where $y$ are the neutrino Yukawa couplings, $M_{N_k}$ are the Majorana masses of the $k=1,2,3$ right-handed neutrinos and $M_{UV}$ is the ultraviolet scale where the supersymmetry breaking parameters originate.\footnote{A different example, where neutrino masses originate from the Type-II seesaw, is discussed in detail \cite{Joaquim:2008bm}.}

Predictions for CLFV processes will depend, in this case, on the values of the neutrino Yukawa couplings, which are not directly constrained by neutrino mass and lepton mixing data. If these are large, expectations are that the rates for several CLFV processes, especially $\mu\to e\gamma$ and $\mu\to e$~conversion in nuclei, are large enough to be seem in current or next-generation experiments. For recent detailed study which discusses different ans\"atze for the neutrino Yukawa couplings see, for example, \cite{Calibbi:2012gr,Blankenburg:2012nx}.

Equation~(\ref{susy_seesaw}) reveals that, at least in principle, one can relate the rates for CLFV processes to a particular combination of neutrino Yukawa couplings. On the other hand, neutrino masses reveal, through the seesaw mechanism, a different combination of neutrino Yukawa couplings: $m_{\nu}\propto \sum_k y_{\alpha k} M_k^{-1} y_{\beta k}$. The hope is that by combining neutrino masses, CLFV and the measurement of super-partner masses and couplings at high energy colliders one can determine the neutrino Yukawa couplings and the right-handed neutrino masses. This would not only nail down the physics behind neutrino masses but would also allow one to test whether the baryon asymmetry of the universe arises due to leptogenesis (see, for example, \cite{Davidson:2008bu} and references therein). Needless to say, in order to test high scale leptogenesis, several ``minor miracles'' have to occur, including: there must be weak scale supersymmetry, and we must measure several supersymmetry breaking parameters very precisely; we must understand the mechanism of supersymmetry breaking reliably; we need to observe several CLFV events in several channels (including some involving tau-leptons); and a few others. For a detailed discussion see, for example, \cite{Buckley:2006nv}. 
 
Other new physics scenarios will lead to different values of $\kappa$, including $\kappa\sim 1$ and $\kappa\gg 1$. Supersymmetric models with trilinear R-parity violation serve as an excellent laboratory for exploring different possibilities \cite{deGouvea:2000cf,Dreiner:2012mx}. In the presence of, say, large couplings $\lambda'$ for the $LQD^c$ superpotential term (see \cite{deGouvea:2000cf} for details and notation), one realizes Eq.~(\ref{l_mec}) in the limit $\kappa\gg 1$ and
\begin{equation}
\frac{1}{\Lambda^2}\sim \frac{\lambda'^2}{M_{\rm SUSY}^2}.
\label{eq:Rpar}
\end{equation} 
In such a scenario the rate for $\mu\to e$~conversion in nuclei is much larger (several orders of magnitude) than that for $\mu\to e\gamma$ and $\mu\to eee$. The presence of other R-parity violating couplings may lead to scenarios where $\kappa\sim 1$ both in Eq.~(\ref{l_mec}) and Eq.~(\ref{eq:l_meee}), while yet another choice leads predominantly to Eq.~(\ref{l_mec}) in the limit $\kappa\gg 1$.

 Large rates for CLFV are not a privilege of supersymmetric scenarios. Any complete new physics scenario that addresses the gauge hierarchy problem will lead to CLFV rates that are much, much larger than those associated to the naive massive neutrino contribution (see Eq.~(\ref{meg_sm})). Concrete examples, associated to different values of $\kappa$, have been calculated for models with large extra dimensions \cite{DeGouvea:2001mz}, Randall-Sundrum models with fermion fields in the bulk \cite{Agashe:2006iy}, little Higgs models \cite{Blanke:2007db}, and many others. 
 
\subsection{CLFV with Tau Leptons}
 
The most stringent bounds on tau-lepton-flavor violation come from rare tau decay processes. Thanks to the B-factories, branching ratios for $\tau\to \ell \gamma$, $\tau\to\ell\ell\ell'$ and $\tau\to\ell +X$, where $\ell,\ell'=e,\mu$ and $X$ are different quark states, are known to be, for the most sensitive decay modes, below a few times $10^{-8}$ \cite{Beringer:1900zz}. Super-B factories, which aim at integrated luminosities larger than 10~ab$^{-1}$, aim at being sensitive to branching ratios larger than $10^{-9}$ (or even a little lower), especially when it comes to ``background-free'' decay modes (e.g. $\tau\to e^+\mu^-\mu^-$).
 
It is not possible to compare, in a model independent way, the sensitivity of rare tau and rare muon processes. At face value, rare muon processes are at least a few orders of magnitude more sensitive than rare tau processes. This remains true when one compares qualitatively similar processes, e.g., $\mu\to e\gamma$ versus $\tau\to e\gamma$, $\mu\to eee$ versus $\tau \to e\mu\mu$, $\mu\to e$~conversion in nuclei versus $\tau\to e\rho^0$, etc. 
 
 There are several new physics scenarios that predict significantly enhanced rare tau process compared to rare muon processes. Enhancement factors depend, of course, on the model. These include chirality effects, which tend to enhance the rate of tau processes by a factor $(m_{\tau}/m_{\mu})^2\sim 280$ and different effects related to neutrino masses or the pattern of lepton mixing. For example, $\tau\to \mu\gamma$ may be enhanced with respect to $\mu\to e\gamma$ by a factor $\Delta m^2_{13}/\Delta m^2_{12}\sim 30$ or  $|U_{\tau 3}/U_{e 3}|^2\sim 20$.
 
\subsection{Final Thoughts}

The discovery of neutrino oscillations reveals that charged-lepton flavor violating phenomena must occur. Naive massive-neutrino expectations are that the rates for CLFV processes are absurdly small thanks to the GIM mechanism and the fact that neutrino masses are tiny compared to the weak scale. Nonetheless, if there are new degrees of freedom at the TeV scale, rates for CLFV processes are expected to be much larger. In most complete scenarios that address the gauge hierarchy problem by introducing new degrees of freedom around the weak scale, expected rates for CLFV are very large, and one is often required to add new conditions or constraints to the new physics sector in order to satisfy bounds from negative searches for CLFV.   

In some new physics scenarios, $\mu\to e\gamma$ is the most promising channel for pursuing searches for CLFV. This, however, is not a universal statement. Indeed, searches for $\mu\to e$~conversion prove to be the ultimate probes of CLFV. Regardless, all searches for CLFV in all channels should be pursued. This is especially true if CLFV is observed in some channel. By combining the results from different probes we will be able to learn much more about the nature of the new physics. 

Finally, CLFV phenomena may be intimately related to the physics behind neutrino masses. They may also play a key role in our understanding of the seesaw mechanism, grand unified theories, and the physics behind the matter--antimatter asymmetry of the universe. Regardless of what new (or old) physics we may discover at the LHC, future studies of CLFV phenomena are guaranteed to provide unique, complementary information -- even if the LHC fails to detect any new degrees of freedom at the TeV scale.

\section{Lepton number conservation}
\label{0nubb}

\subsection{Origin of the neutrino mass}

Neutrino masses are  six or more orders of
magnitude smaller than the masses of other charged fermions. Moreover,
the pattern of masses, i.e. the mass ratios of neutrinos, is rather different
(even though it remains largely unknown, one does not know whether it follows
the normal or inverted hierarchy or whether the pattern of masses
is essentially degenerate ) than the pattern of masses
of the up- or down-type quarks or charged leptons. And the neutrino mixing
matrix does not look at all as its analog, the CKM mixing matrix of the down-type
quarks. All of these facts
suggest that, perhaps, the origin of the neutrino mass is different
than the origin (which is still not well understood  but is believed to be the coupling
to the Higgs vacuum expectation value) of the masses of the other
charged fermions.

The smallness of the neutrino masses can be possibly understood following the
finding of Weinberg \cite{Wei79} who pointed out more than thirty year ago
that there exists only one
lowest order (dimension 5, suppressed by only one inverse power of the
corresponding high energy scale $\Lambda$) gauge-invariant operator given the content
of the Standard Model. After the spontaneous symmetry
breaking when the Higgs acquires vacuum expectation value that operator
represents the neutrino Majorana mass which violates the total lepton number
conservation law by two units,
\begin{equation}
{\cal L}^{(M)} = \frac{C^{(5)}}{\Lambda} \frac{ v^2}{2} (\bar{\nu}^c \nu) + h.c. ~~,
\end{equation}
where $v \sim$ 250 GeV and $C^{(5)}$ might be of the order of unity.
For sufficiently large scale $\Lambda$ neutrino masses are arbitrarily small.
There exist a large number of physics realization of this basic idea. They all
necessarily lead to the conclusion that neutrinos are Majorana fermions and
that the total lepton number is not exactly conserved quantity. 

The most popular physics model
explanation of the smallness of neutrino mass is the seesaw mechanism,
which is also roughly thirty years old \cite{seesaw1,seesaw2}. In it, the existence of heavy right-handed
neutrinos $N_R$ is postulated, and by diagonalizing the corresponding mass matrix
one arrives at the formula
\begin{equation}
m_{\nu} = \frac{m_D^2}{M_N}
\end{equation}
where the Dirac mass $m_D$ is assumed to be a typical charged fermion mass and $M_N$ is
the Majorana mass of the heavy neutrinos $N_R$. Again, the small mass of the standard neutrino
is related to the large mass of the heavy right-handed partner. Requiring that $m_{\nu}$ is of the
order of 0.1 eV means that $M_N$ (or $\Lambda$) is $\sim 10^{14-15}$ GeV, i.e. near the GUT
scale. That makes this, so-called Type I seesaw template scenario particularly attractive. 
However, clearly an observation of the gauge-singlet neutrinos $N_R$, if they exist, is
not possible.

There are, however, other possible scenarios that lead to a small neutrino Majorana mass.
In the Type II seesaw one adds a Higgs triplet $(\xi^{++}, \xi^+, \xi^0 )$ which couples
directly to the symmetric triplet combination of two $(l, \nu)_L$ doublets. In that case
\begin{equation}
m_{\nu} = h_{\nu} \langle \xi^0 \rangle ~,
\end{equation}
that could be small provided that the $ \langle \xi^0 \rangle $ is very small, even for a
``natural'' value of the Yukawa coupling $h_{\nu}$
(see the discussion following Eq. (\ref{eq:ssII}) in Section IID). 
It works since the spontaneous breaking 
of electroweak symmetry is accomplished by the vacuum expectation value of 
$\langle \Phi^0 \rangle = v$,
of the ordinary Higgs and thus  $ \langle \xi^0 \rangle $ could be small.
If the mass of the $\xi^{++}$, $M_{\xi} \sim$ 1 TeV then its decay into $l^+_i l^+_j$ could be observable
and would be another signal of  the lepton number violation.
In Type III the singlet neutrinos $N_R$ are replaced by a fermion triplet
$\Sigma^+, \Sigma^0, \Sigma^-$. Small neutrino Majorana masses are again obtained. 
(See Ref. \cite{Ma} for a pedagogical treatment of the seesaw mechanism.)  

\begin{figure}[tb]
\begin{center}
\begin{minipage}[t]{8cm}
\centerline{\epsfig{file=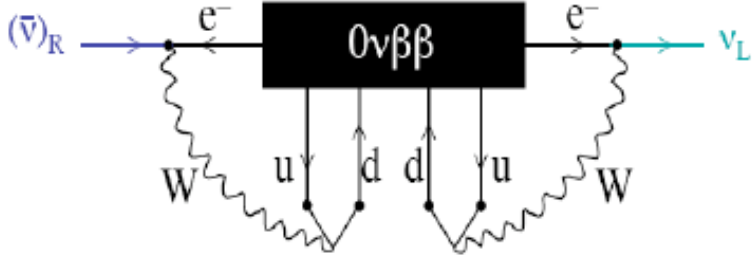,scale=0.4}}
\end{minipage}
\begin{minipage}[t]{16.5cm}
\caption{ By adding loops involving only standard weak interaction processes the
$0\nu\beta\beta$ decay amplitude (the black box)
implies the existence of the Majorana neutrino mass \cite{SV}.}
\label{fig:SV}
\end{minipage}
\end{center}
\end{figure}

All of these possibilities not only offer an explanation of the smallness of neutrino masses,
but they also suggest that neutrinos are likely massive Majorana fermions. To test such
hypothesis one should explore various ways of lepton number violation.
In that context it is important to point out, as shown by Schechter and Valle \cite{SV},
that the observation of lepton number violating process, in the case considered 
by them,  the neutrinoless
double beta decay $dd \rightarrow uuee$, 
implies that neutrinos are massive Majorana particles
(see Fig. \ref{fig:SV} for illustration). This theorem, however,
does not mean that it is easy (or even generally possible) to determine the neutrino
mass once the $0\nu \beta\beta$ decay is observed and its decay rate determined.
In fact, the multiloop graph in Fig.~\ref{fig:SV} only illustrates the fundamental relation
between the $0\nu\beta\beta$ decay and the Majorana neutrino mass. Taking into account
the upper limit for the  $0\nu\beta\beta$ amplitude, deduced from the lower limit
of the  $0\nu\beta\beta$ decay half-life, one comes to the conclusion that the Feynman graph
in Fig.~\ref{fig:SV} corresponds to an unobservably small neutrino Majorana mass \cite{Lindner}.
The true origin of the neutrino masses must be elsewhere, probably in one
of the possible tree level diagrams.
Nevertheless the graph in Fig.~\ref{fig:SV}  illustrates the intimate relation between the 
neutrino Majorana mass on one hand, and the neutrinoless double beta decay on the other hand. 

\subsection{Testing the total lepton number}

As stresses in the Introduction, testing whether the total lepton number is conserved,
i.e. whether $U(1)_L$ is a good symmetry or not, is of fundamental importance.
It separates the form of the ``New Standard Model" into two distinct classes with massive
Dirac or Majorana neutrinos. 

There are various ways to test whether the total lepton number conservation. No violation
has been seen so far.
Examples of the potentially lepton number violating (LNV)
processes with important limits are
\begin{eqnarray}
&& (Z,A) \rightarrow (Z+2,A) + 2e^-; {~\rm half-life~ > ~10^{25} ~ years}
\nonumber \\
&& \mu^- + (Z,A)  \rightarrow  e^+  + (Z-2,A); {~\rm exp.~ branching~ ratio} \le 10^{-12} ~,
\nonumber \\
&&  K^+  \rightarrow   \mu^+ \mu^+ \pi^-;  {~\rm exp. ~branching ~ratio} \le 3 \times 10^{-9} ~,
  \nonumber \\
&&  \bar{\nu}_e {\rm ~emission~ from~ the~ Sun};  {~\rm exp.~ branching~ ratio} \le 10^{-4} ~.
\end{eqnarray}

Detailed analysis suggests that the study of the $0\nu\beta\beta$ decay,
the first on the list above, is by far the
most sensitive test of LNV. In simple terms this is caused by the amount of tries one can make.
A 100 kg $0\nu\beta\beta$ decay source contains $\sim 10^{27}$ nuclei
that can be observed for a long time (several years). This can be contrasted
with the possibilities of first producing muons or kaons, and then searching for the unusual
decay channels. The Fermilab accelerators, for example, produce $\sim 10^{20}$ protons on target
per year in their beams and thus correspondingly smaller numbers of muons or kaons.
However, the statement about the overwhelming sensitivity to LNV of the $0\nu\beta\beta$ decay  
is based on the assumption that the source of the lepton number nonconservation
is the fact that the light neutrino mass eigenstates $\nu_1, \nu_2$ and $\nu_3$ are selfconjugate
Majorana particles. It also assumes that there is no symmetry that would cause the
corresponding $m_{ee}$ entry of the neutrino mass matrix in the flavor basis to vanish.

However, the origin of the lepton number nonconservation could be also elsewhere. It can be 
related to the existence of $\sim$ TeV mass,  
so far unobserved particles, e.g. a double charged Higgs
$\xi^{++}$ of the seesaw Type II scenario,
that can directly decay into a pair of the same charge leptons. If one could produce a sufficient number
of such particles, and detect their decay, that would represent another test of the lepton
number conservation. Examples of the particle physics models of such type and estimates of
the corresponding production cross sections could be found e.g. in Ref.\cite{Pavel} where other
references for analogous ideas could be found.
Some of these processes could, in principle, lead to the observation of LNV even given the existing 
limits from the  $0\nu\beta\beta$ decay. 

Nevertheless, in the following we shall discuss the search for LNV using the $0\nu\beta\beta$ decay
as the probe, and the corresponding physics involved.

\subsection{Brief history of $0\nu\beta\beta$ decay searches}

In double beta decay two neutrons, bound in the ground state of an even-even initial 
(or parent) nucleus
are transformed into two bound protons, typically also in the the ground state of the
final (or granddaughter) even-even nucleus, with the simultaneous emission of two
electrons only for the $0\nu\beta\beta$ mode, or two electrons plus two $\bar{\nu}_e$
for the $2\nu\beta\beta$ mode
\begin{equation}
(Z,A)_{g.s.} \rightarrow (Z+2,A)_{g.s.} + 2e^- + (2\bar{\nu}_e) ~.
\label{eq:1}
\end{equation}
The $\beta\beta$ decay, in either mode, can proceed only if the initial nucleus is stable against
the standard $\beta$ decay (both $\beta^-$ and $\beta^+$ or $EC$). That happens exclusively
in even-even nuclei where, moreover, the ground state is always $I^{\pi} = 0^+$.
The $\beta\beta$ decay rate is a steep function of the energy carried by the 
outgoing leptons (i.e. of the decay $Q$-value). Hence, transitions with larger $Q$-value
are easier to observe and experimental search is centered on such nuclei.
Possible exception to this rule is the resonant neutrino less double electron capture
($0\nu ECEC$) where $Q \rightarrow 0$. This topic will not be further discussed
here, but its theory is comprehensively covered in, e.g. Ref. \cite{SimEC}.

In both modes of the  $\beta\beta$ decay the rate can be expressed as a product
of independent factors that depend on the atomic physics (the so called phase-space 
factors $G^{0\nu}$ and $G^{2\nu}$) that include
also the $Q$-value dependence as well as the fundamental physics constants which
can be evaluated accurately, nuclear structure (the nuclear
matrix elements $M^{0\nu}$ and $M^{2\nu}$) that
need to be evaluated using the nuclear structure theory, and thus are known 
but only with considerable uncertainty, 
and for the $0\nu\beta\beta$ mode the possible
particle physics parameters (the effective neutrino mass $\langle m_{\beta\beta} \rangle$ in the
simplest case). These particle physics parameters are, naturally, the important output
of the search for the $0\nu\beta\beta$ decay. Thus
\begin{equation}
\frac{1}{T_{1/2}^{0\nu}} = G^{0\nu} |M^{0\nu}|^2 |\langle m_{\beta\beta} \rangle|^2 ~;
\hspace{1cm} \frac{1}{T_{1/2}^{2\nu}} =  G^{2\nu} |M^{2\nu}|^2 ~.
\label{eq:rate}
\end{equation} 

\begin{figure}[tb]
\begin{center}
\begin{minipage}[t]{8cm}
\centerline{\epsfig{file=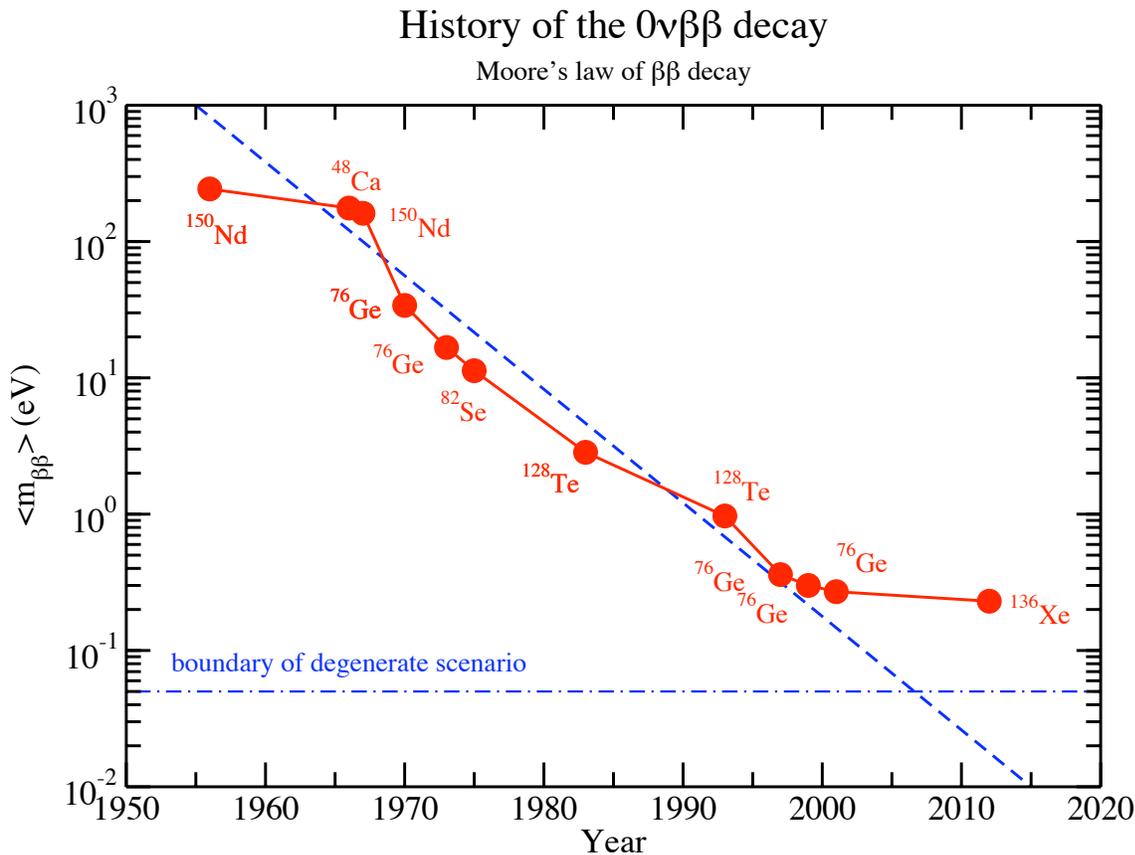,scale=0.6}}
\end{minipage}
\begin{minipage}[t]{16.5cm}
\caption{ Moore's law of  $0\nu\beta\beta$ searches. The past experiments are characterized by the
sensitivity to the effective neutrino Majorana mass $\langle m_{\beta\beta} \rangle$.}
\label{fig:history}
\end{minipage}
\end{center}
\end{figure}

The phase space factors $G^{0\nu}$ and $G^{2\nu}$ can be found in \cite{B-V}. More recent,
somewhat different values of them that include effects of the nuclear size, are published in \cite{K-I}.

The search for the $0\nu\beta\beta$ decay has a very long history. There have been over hundred
published results in the last fifty years. The most sensitive ones, expressed as the upper limits
of the effective Majorana mass $<m_{\beta\beta}>$ are shown in Fig.~\ref{fig:history}. Over most of 
that time the trend of the time dependence more or less followed a straight line on the 
semilog plot as indicated by the blue dashed line (the Moore's law of $\beta\beta$ decay).
That led to the improvement in sensitivity to  $<m_{\beta\beta}>$ by about an order of magnitude
per decade (i.e. by a factor of $\sim$100 in half-life). 
However, it is a bit worrisome to note that during the last decade that trend was no longer
followed, and only slight improvement was achieved very recently. This shows that the
experiments have become qualitatively more complex and costly; it might be challenging 
to maintain the historic trend in near future.

\begin{figure}[htb]
\begin{center}
\begin{minipage}[t]{8cm}
\centerline{\epsfig{file=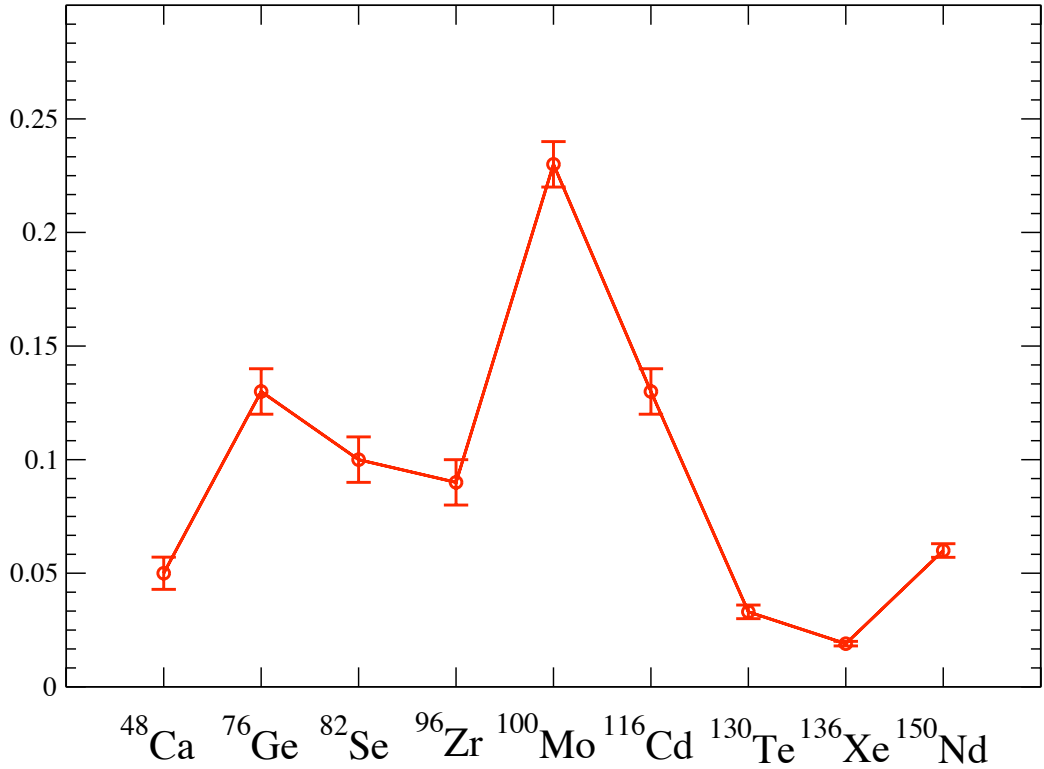,scale=0.6}}
\end{minipage}
\begin{minipage}[t]{16.5 cm}
\caption{ Matrix elements $M^{2\nu}$ in MeV$^{-1}$ based on the
experimental halflife measurements. }
\label{fig_2nu}
\end{minipage}
\end{center}
\end{figure}

At the same time, during the last two decades the less fundamental 
$2\nu\beta\beta$ decay has been observed
in ``live" laboratory experiments
in many nuclei, often by different groups and using different
methods. That shows not only the ingenuity of the experimentalists who
were able to overcome the background nemesis, but makes it possible
at the same time to extract the corresponding $2\nu$ nuclear matrix element
from the measured decay rate. 
Study of the $2\nu\beta\beta$ decay is an interesting nuclear physics 
problem by itself. Moreover, evaluation of the $M^{2\nu}$ matrix elements
is an important test for the nuclear theory models that aim at the determination
of the analogous but different quantities for the more fundamental $0\nu$ neutrinoless mode.
(We will discuss that issue later.)
The resulting nuclear matrix elements $M^{2\nu}$, 
which have the dimension energy$^{-1}$,
are plotted in Fig.~\ref{fig_2nu}. (The corresponding experimental half-lives
$T_{1/2}$ were taken from the Refs. \cite{Balysh, Doerr,Arnold, Argyr,
Danev,Lin,Arn11,Acker,Argyr2,Turk}.)
Note the pronounced shell dependence; the matrix element
for $^{100}$Mo is almost ten times larger than the ones for $^{130}$Te or $^{136}$Xe.

 \subsection{Mechanism of the $0\nu\beta\beta$ decay}
 
The relation between the $0\nu\beta\beta$-decay
rate and the effective Majorana mass $\langle m_{\beta\beta} \rangle$
is to some extent problematic. The rather conservative
assumption leading to Eq.(\ref{eq:rate}) is that, if $0\nu\beta\beta$ decay
will occur at an observable rate, it will proceed dominantly through the exchange
of a virtual light, but massive,
Majorana neutrino between the two nucleons undergoing the transition,
and that these neutrinos interact by the standard left-handed weak currents. But that is not
the only theoretically possible mechanism. 
Lepton number violating (LNV) interactions involving so far unobserved
much heavier ($\sim$ TeV) particles
might potentially lead to a comparable $0\nu\beta\beta$ decay rate.

In general $0\nu\beta\beta$ decay can be generated by (i) light massive Majorana
neutrino exchange, as in the Eq.(\ref{eq:rate}) or (ii) heavy particle exchange (see, e.g. Refs.\cite{heavy,Pre03}),
resulting from LNV dynamics at some scale $\Lambda$ above the electroweak one.
(That can happen e.g. in the seesaw Type II and III scenarios.)
The relative size of heavy ($A_H$) versus light
particle ($A_L$) exchange contributions to the decay amplitude
can be crudely estimated as follows~\cite{Mohapatra:1998ye}:
\begin{equation}
A_L \sim G_F^2  \frac{\langle m_{\beta \beta} \rangle}{\langle k^2 \rangle}  ,~
 A_H \sim G_F^2  \frac{M_W^4}{\Lambda^5}  ,~
\frac{A_H}{A_L} \sim \frac{M_W^4 \langle k^2 \rangle }
{\Lambda^5  \langle m_{\beta \beta} \rangle }  \ ,
\label{eq_estimate}
\end{equation}
where $\langle m_{\beta \beta} \rangle$ is the effective neutrino
Majorana mass,
$\langle k^2 \rangle \sim ( 100 \ {\rm MeV} )^2 $ is the
typical light neutrino virtuality, and $\Lambda$ is the heavy
scale relevant to the LNV dynamics.
Therefore,  $A_H/A_L \sim O(1)$ for  $\langle m_{\beta \beta} \rangle \sim 0.1-0.5$
eV and $\Lambda \sim 1$ TeV, and  thus the LNV dynamics at the TeV
scale would lead to similar $0 \nu \beta \beta$-decay rate as the
exchange of light Majorana neutrinos with the effective mass
$\langle m_{\beta \beta} \rangle \sim 0.1-0.5$ eV.

Obviously, the $0\nu\beta\beta$ lifetime measurement by itself
does not provide the means for determining the underlying mechanism.
The spin-flip and non-flip exchange can be, at least in principle,
distinguished by the, naturally very difficult, measurements 
of the single-electron spectra or of the electron
polarizations.  However, in most cases the
mechanism of light Majorana neutrino exchange, and of
heavy particle exchange, cannot be separated by the observation
of the emitted electrons. 

The main difference between the mechanism 
involving the light massive Majorana neutrino exchange and the heavy ($\sim$ TeV)
particle exchange is the range of the operator that causes the transition. The light neutrino
exchange represents two point-like vertices separated by the distance $r \sim 1/q$,
with $q \sim O(100)$ MeV.
The decay rate is then proportional to the square of the effective Majorana
neutrino mass as in Eq. (\ref{eq:rate}). On the other hand
the heavy particle exchange represents a single point-like vertex
(six fermions, four hadrons and two leptons), i.e. dimension 9 operator. 
Proper treatment of the short range nucleon-nucleon repulsion and the effects
related to the finite size of the nucleon is obviously crucial
in that case. The relation between the
neutrino mass and the decay rate is very indirect in that case, but extraction of various
important particle physics parameters is possible when the corresponding particle
physics model has been specified. 

\subsection{Specific models: LRSM and RPV-SUSY}

Lets consider the case where the scale of LNV is relatively low ($\sim$ TeV).
The relevant particle physics models in that case usually contain not only 
the possibility of Lepton Number Violation but also a possibility of the 
Lepton Flavor Violation (CLFV) involving charged leptons. 
Denoting the new physics scale by $\Lambda$, one has a LNV
effective lagrangian of the form
\begin{equation}
{\cal L}_{0 \nu \beta \beta} = \displaystyle\sum_i \
\frac{\tilde{c}_i}{\Lambda^5}  \  \tilde{O}_i
\qquad  \tilde{O}_{i} =  \bar{q} \Gamma_1 q \,  \
\bar{q} \Gamma_2 q \,   \bar{e} \Gamma_3 e^c   \ ,
\label{eq:lag1}
\end{equation}
where we have suppressed the flavor and Dirac structures
(a complete list of the dimension nine operators
$\tilde{O}_i$ can be found in Ref.~\cite{Pre03}).

For the CLFV interactions, one has operators of dimension six,
\begin{equation}
{\cal L}_{\rm CLFV} = \displaystyle\sum_i \
\frac{c_i}{\Lambda^2}  \  O_i  \  ,
\label{eq:lag2}
\end{equation}
The CLFV operators relevant to
our analysis are of the following type (along with their analogues
with $L \leftrightarrow R$):
\begin{equation}
O_{\sigma L}  =    \displaystyle\frac{e}{(4 \pi)^2}
 \overline{\ell_{iL}} \, \sigma_{\mu \nu} i
/ \hspace{-0.23cm}D \, \ell_{jL}  \  F^{\mu \nu}  + {\rm h.c.} ~,
~~O_{\ell L}  =   \overline{\ell_{iL}} \, \ell^c_{jL} \
\overline{\ell^c_{kL}} \, \ell_{mL}~,
~~O_{\ell q}  =    \overline{\ell_{i}} \Gamma_\ell \ell_{j} \
\overline{q} \Gamma_q  q  \  .
\end{equation}

Operators of the type $O_{\sigma}$ are typically generated at one-loop level,
hence our choice to explicitly display the loop factor
$1/(4 \pi)^2$.   On the
other hand, in a large class of models, operators of the type $O_{\ell}$ or
$O_{\ell q}$ are generated by tree level exchange of heavy degrees of
freedom. With the above choices,
all non-zero $c_i$ are nominally of the same size,
typically the product of two Yukawa-like couplings or gauge couplings
(times flavor mixing matrices). These operators are analogous
to the operators in Eqs. (\ref{l_mec}) and (\ref{eq:l_meee}) but 
normalized differently in order to follow the notation of Ref.\cite{Vinc}.

Whenever the operators $O_{\ell L,R}$ and/or $O_{\ell q}$ appear at
tree-level in the effective theory, they lead to an enhancement factor
$\log \left( \frac{\Lambda^2}{m_\mu^2} \right) $ in the LFV processes 
involving charged muons $\mu \rightarrow e$
(muon conversion) or $\mu \rightarrow eee$. That logarithmic enhancement 
is absent in the $\mu \rightarrow e\gamma$ process.

Based on our analysis in Ref. \cite{Vinc} we can formulate the following approximate
rules ( a diagnostic tool) regarding the relative size of the branching ratios 
for  $\mu \rightarrow e$ and $\mu \rightarrow e\gamma$ and the relation of that
size to the mechanism of the $0\nu\beta\beta$ decay:
\begin{enumerate} 
\item 
Observation of both the CLFV muon processes
$\mu \to e$ and $\mu \to e \gamma$ with relative branching ratios $\sim
10^{-2}$ implies, under generic conditions, that $\Gamma_{0 \nu \beta
\beta} \sim \langle m_{\beta \beta} \rangle^2$. Hence the relation
of the $0\nu\beta\beta$ lifetime to the absolute neutrino mass scale
is straightforward.
\item 
On the other hand, observation of CLFV muon processes with
relative branching ratios $ \gg 10^{-2}$ (presumably caused by the log enhancement
factor $\frac{\Lambda^2}{m_\mu^2} $ ) could signal non-trivial LNV
dynamics at the TeV scale, whose effect on $0 \nu \beta \beta$ has to
be analyzed on a case by case basis. Therefore, in this scenario no
definite conclusion can be drawn based on LFV rates.
\item
Non-observation of CLFV in muon processes in forthcoming substantially more sensitive
experiments would imply either that the scale of non-trivial CLFV and
LNV is  above a few TeV, and thus $\Gamma_{0 \nu \beta
\beta} \sim \langle m_{\beta \beta} \rangle^2$, or that any TeV-scale LNV is
approximately flavor diagonal.
\end{enumerate}

The above statements were illustrated using two explicit cases: the
Minimal Supersymmetric Standard Model  with R-parity violation
(RPV-SUSY) and the Left-Right Symmetric Model (LRSM) in Ref. \cite{Vinc}.

It is likely that the basic mechanism at work in 
these illustrative cases is  generic: low scale LNV interactions
($\Delta L = \pm 1$ and/or $\Delta L= \pm 2$), which in general
contribute to $0 \nu \beta \beta$, also generate sizable contributions
to $\mu \to e$ conversion, thus enhancing this process over $\mu \to e
\gamma$ (see also discussion following Eq. (\ref{eq:Rpar}) in  Section IID.

\subsection{$0\nu\beta\beta$ nuclear matrix elements}

Double beta decay in both $2\nu$ and $0\nu$ modes can exist because even-even nuclei
are more bound than the neighboring odd-odd nuclei. This extra binding
is a consequence of {\it pairing} between like nucleons. In nonmagic systems
neutrons and/or protons form $0^+$ pairs and the corresponding Fermi level becomes
diffuse over the region  with the characteristic size $\sim$ pairing gap $\Delta$.
This opens more possibilities for $nn \rightarrow pp$ transitions. The calculated
matrix elements $M^{0\nu}$ increase when the gap $\Delta$ increases. The 
(unrealistic in real nuclei) situation of pure paired nuclear system (only seniority 0
states) would have very large $M^{0\nu}$.

However, in real nuclei opposite tendencies are also present. Real nuclei have admixtures of 
the ``broken pair'' states, or in the shell model language, states with higher
seniority. These states are present because other parts of the nucleon interaction
exist, in particular the neutron-proton force. It is illustrative to characterize 
such states by the angular momentum $\cal{J}$ of the neutron pair that is in the
$\beta\beta$ decay transformed into the proton pair with the same $\cal{J}$.
While the pairing parts  $\cal{J} = $ 0 are large and positive, the $\cal{J} \ne$ 0 parts are negative
and their sum is essentially as large as the $\cal{J} = $ 0 piece. 
The severe cancellation between these
two tendencies, as a consequence of the corresponding components of the residual interaction,
is present in most methods of evaluating the $M^{0\nu}$. That makes their accurate 
determination challenging.

\begin{figure}[tb]
\begin{center}
\begin{minipage}[t]{8 cm}
\centerline{\epsfig{file=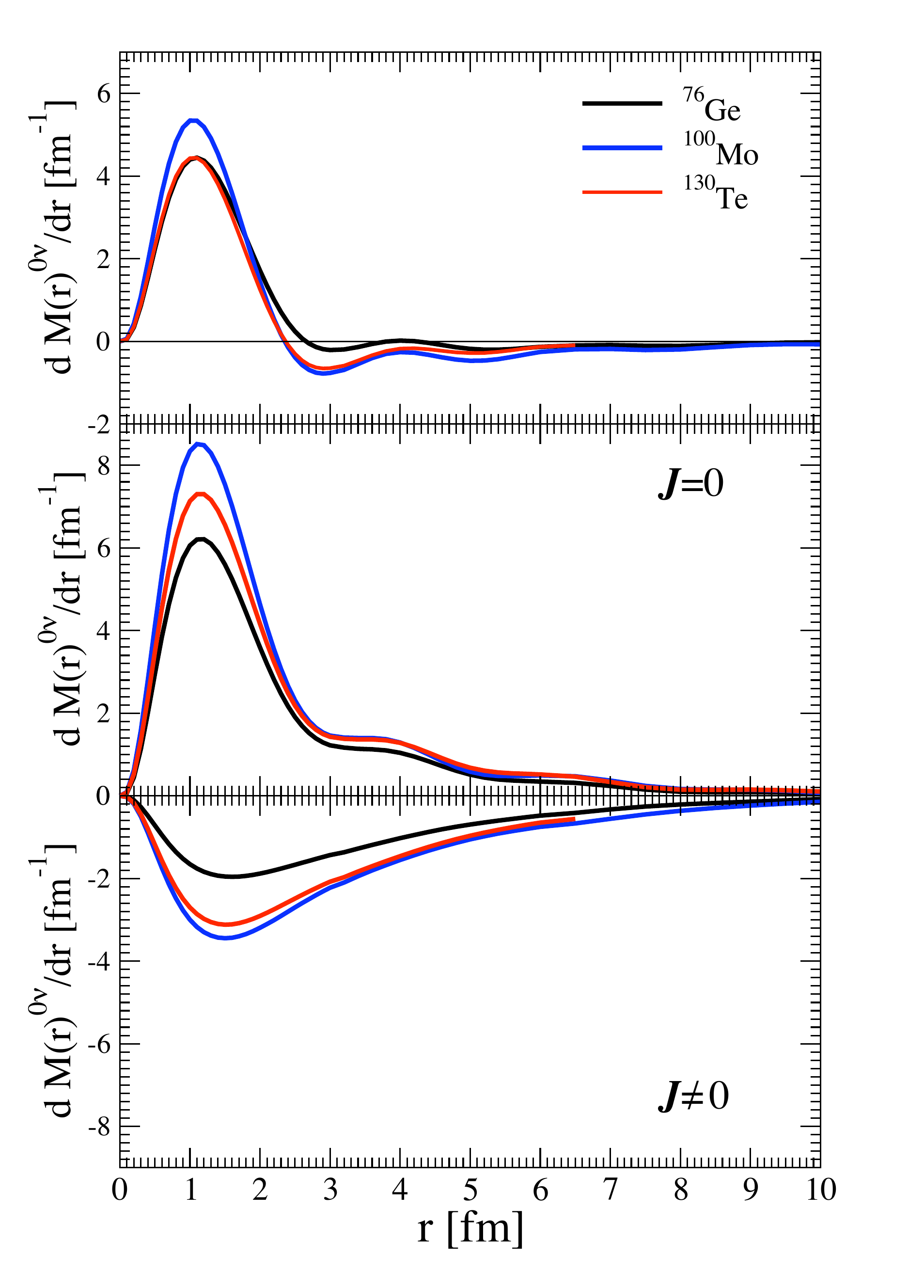,scale=0.5}}
\end{minipage}
\begin{minipage}[t]{16.5 cm}
\caption{The dependence on $r_{12}$ for $^{76}$Ge, $^{100}$Mo and $^{130}$Te
evaluated in QRPA. The upper panel shows the full matrix element, and the lower
panel shows separately the ``pairing'' ($\cal{J}$ = 0) and ``broken pair''
 ($\cal{J} \ne$ 0) contributions. The integrated $M^{0\nu}$ are 5.4, 4.5, and 4.1
for the three nuclei. }
\label{fig:radial}
\end{minipage}
\end{center}
\end{figure}

The $0\nu\beta\beta$ decay operators depend on the internucleon distance $r_{12}$ 
due to the neutrino potential $H(r,\bar{E})$
(which is essentially the Fourier transform of the light neutrino propagator). 
Obviously, the range of $r_{12}$ is restricted
from above by $r_{12} \le 2R$. From the  form of $H(r) \sim R/r$ one could, naively,
expect that the characteristic value of $r_{12}$ is the typical distance between nucleons
in the nucleus, namely $\bar{r}_{12} \sim R$. However, that is not true, in reality only
much smaller values of $r_{12} \le$ 2-3 fm or equivalently larger values of the
momentum transfer $q$ are relevant. 
The qualitatively same conclusion was reached in both widely used approximate
methods, the Quasiparticle Random Phase Approximation (QRPA) \cite{anatomy}
and in the Nuclear Shell Model (NSM) \cite {Men09}. Examples of the functions 
$C^{0\nu}(r)$ are shown in Fig. \ref{fig:radial} for
three representative nuclei. As the lower panel demonstrates, the cancellation
between the ``pairing'' ($\cal{J}$ = 0) and ``broken pairs'' ($\cal{J} \ne$ 0) is 
essentially complete for $r_{12} \ge$ 2-3 fm

To see how that conclusion is obtained, we define the function $C(r)$
\begin{equation}
C^{0\nu}_{GT}(r) =  \langle f | \Sigma_{lk} \vec{\sigma}_l \cdot  \vec{\sigma}_k \tau_l^+ \tau_k^+
\delta(r - r_{lk}) H(r_{lk},\bar{E}) | i \rangle ~,
\label{eq:C(r)}
\end{equation}
Obviously, this function is normalized by
\begin{equation}
 M^{0\nu}_{GT} = \int_0^{\infty} C^{0\nu}_{GT} (r) dr ~.
 \label{eq:C(r)int}
\end{equation}
The definition above is valid  for the  Gamow-Teller part of $M^{0\nu}$, which is, 
however, its most important part. In 
Fig.~\ref{fig:radial} the full $C^{0\nu}(r)$ is plotted (analogous figure appeared in \cite{anatomy}). 
Note that due to the relatively short
range of  $C^{0\nu}(r)$ proper treatment of the nucleon short-range repulsion, nucleon 
finite size, as well as of the recoil order parts of the nucleon weak current (induced pseudoscalar,
weak magnetism) are necessary.

Variety of methods has been used for evaluation of the $0\nu\beta\beta$
nuclear matrix elements for the simplest scenario where the decay rate is proportional
$< m_{\beta\beta} >^2$. They differ in their choice of the valence space,
interaction hamiltonian and the ways the corresponding equations of motion 
are solved.  Let us stress that
an exact {\it ``ab initio}'', i.e. without approximations, calculation of $M^{0\nu}$ for the
candidate nuclei is impossible at the present time.  

The nuclear shell model (NSM) is, in principle, the method that seems to be
well suited for this task. In it, the valence space consists of just few
single particle states near the Fermi level. 
With interaction that is based on the realistic nucleon-nucleon force,
but renormalized slightly to describe better masses, energies and transition
probabilities
in real nuclei, all possible configurations of the valence nucleons
are included in the calculation. 
The resulting states have not only the correct number of protons and neutrons, 
but also all relevant quantum numbers (angular momentum, isospin, etc). 

The quasiparticle random phase approximation (QRPA) and its renormalized version
(RQRPA) is another method often used in the evaluation of $M^{0\nu}$. In it,
the valence space is not restricted and contains at least two full oscillator
shells, often more than that. On the other hand, only selected 
simple configurations of the valence nucleons are used. The basis states have
broken symmetries in which particle numbers, isospin, and possibly angular
momentum are not good quantum numbers but conserved only on average. 
After the equations of motion are solved, some of the symmetries are partially restored. The RQRPA
partially restores the Pauli principle violation in the resulting states.
Thus, in certain sense the NSM and QRPA are complementary methods.

The IBM-2 method uses the microscopic interacting boson model to
evaluate $M^{0\nu}$ \cite{Barea09}. In IBM-2 one begins with 
correlated $S$ (angular momentum 0) and $D$
(angular momentum 2) pairs of identical nucleons and includes
the effect of deformation through the bosonic neutron-proton
quadrupole interaction. The method describes well the low lying
states, the electromagnetic transitions between them and the two-nucleon
transition rates in spherical and strongly deformed nuclei. 
Even though the method was originally considered as an
approximation of the nuclear shell model, the resulting $M^{0\nu}$
are, rather
surprisingly, close to the RQRPA results and noticeably larger than the NSM ones.
At the present time it is not possible to evaluate the $M^{2\nu}$ matrix
elements within the IBM-2 method, except in the closure approximation.

In Ref. \cite{Rod10} the generating coordinate method (GCM or EDF) was employed.
Large single particle space was used (11 shells) with the well established Gogny D1S
energy density functional. The initial and final many-body wave
functions are represented as combinations of the  particle number $N,Z$ projected,
$I=0^+$ axially symmetric states with different intrinsic deformations.  Again, like in IBM-2
method, the treatment of the odd-odd nuclei, and thus also of the $M^{2\nu}$
matrix elements is impossible in GCM. 

Finally, the PHFB method, used in Ref. \cite{PHFB}, uses the projected 
Hartree-Fock-Bogolyubov wave function. 

\begin{figure}[tb]
\begin{center}
\begin{minipage}[t]{8 cm}
\centerline{\epsfig{file=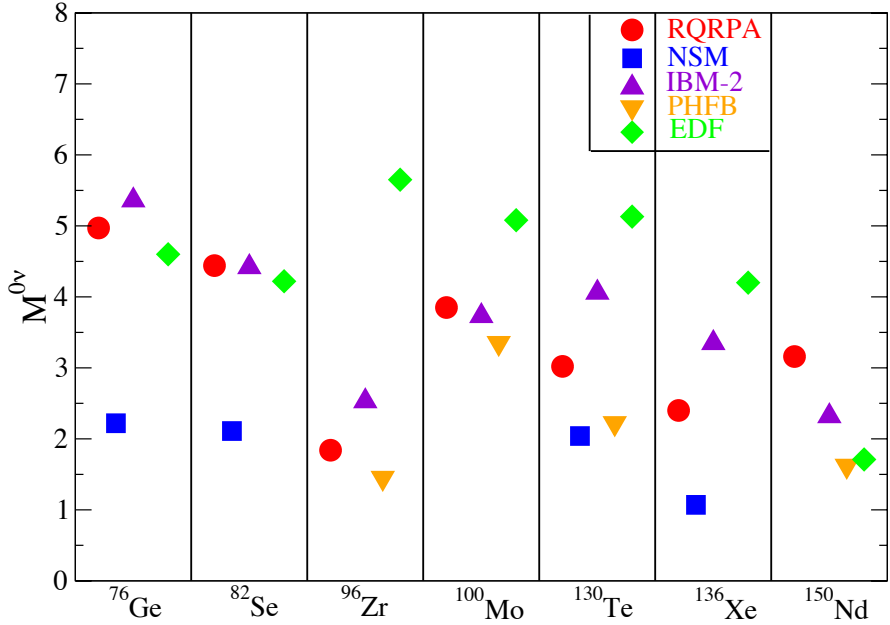,scale=0.5}}
\end{minipage}
\begin{minipage}[t]{16.5 cm}
\caption{Dimensionless $0\nu\beta\beta$ nuclear matrix elements for selected nuclei
evaluated using a variety of indicated methods. For references see text.}
\label{fig:nme}
\end{minipage}
\end{center}
\end{figure}

The results of all of these five methods for the most important candidate nuclei
are shown in Fig.~\ref{fig:nme} (NSM is from \cite{Men09}, IMB-2 from \cite{Barea09},
EDF from \cite{Rod10}, RQRPA from \cite{Sim09} and \cite{Fang10})
and PHFB from \cite{PHFB}.
There one can see that, common to all  displayed
methods, the predicted $M^{0\nu}$ nuclear matrix elements vary relatively smoothly,
with the mass number $A$, unlike the experimentally determined $M^{2\nu}$ matrix elements
displayed in Fig. \ref{fig_2nu} that vary strongly with $A$. 
The RQRPA, IBM-2 and GCM methods are in a crude 
agreement with each other, and predict slow decrease of $M^{0\nu}$ with increasing $A$.
The $M^{0\nu}$ evaluated in NSM are essentially constant with $A$ and noticeably
smaller than those from the other methods, particularly in the lighter nuclei.

As stressed earlier, when heavy particles of any kind mediate the $0\nu\beta\beta$ decay, 
we are dealing with a six fermion vertex, representing extremely short range operator. 
Traditionally, the short range nucleon-nucleon repulsion, which would heavily suppress the
matrix elements  in that case, is  overcome by introducing  the dipole type nucleon form factor (see
\cite{Ver81}). The neutrino potential will be then of the form
\begin{equation}
H_{heavy\nu}(r, M_A) = \frac{4\pi R}{M_A^2} \int \frac{d \vec{q}}{(2\pi)^3} \left( 
\frac{M_A^2}{M_A^2 + \vec{q}^2} \right)^4 = \frac{M_A R}{48} e^{-M_A r} 
\left[ 1 + M_A r + \frac{1}{3} (M_a r)^2 \right]
\end{equation}
Such potential will have the range $1/M_A \sim$ 1/GeV and will be much less affected by 
the short range correlations. The disadvantage of the form factor modeling is that the error introduced
by such approximation is very difficult to estimate.

However, as pointed out in Ref. \cite{Fae97} and analyzed in detail 
in \cite{garybb} using the Effective Field Theory (EFT)
approach, the more effective and fundamentally leading effect is to 
replace the $NNNNee$ six fermion vertex by the $\pi\pi ee$ vertex.
Concrete application of the pion exchange mechanism
suggests the dominance of the $\pi\pi ee$ over the
short range nucleon only vertex by a factor of 10 - 30 at the level of the
nuclear matrix element. This remains, to some extent, so far largely unexplored
issue. 

\subsection{Possible existence of light sterile neutrinos}

Most particle physics models of neutrino mass contain additional, so far unobserved, gauge
singlet neutral fermions, commonly identified as sterile or right-handed neutrinos. Such $\nu_R$ could 
be very heavy, such as in
the orthodox Type I seesaw, and hence unobservable. In other models 
(e.g. in seesaw Type II and III)
such sterile neutrinos have $\sim$TeV masses, and their mixing with the three active neutrinos
is again expected to be very small. Existence of sterile neutrinos in models of this type is needed
and not surprising;  without them proper explanation of the masses of the observed light 
active neutrinos would be impossible.

Remarkably, recently there appeared several  empirical indications, so far of somewhat limited statistical 
significance and mutual consistency,
that very light (of mass $O(1 eV)$) sterile neutrinos might exist, 
that moreover mix noticeably with the flavor neutrinos.
Since these hints appear in a variety of independent experiments, it is worthwhile to analyze
their possible effect on the problem discussed here, namely on the $0\nu\beta\beta$ decay.
One has to keep in mind, however, that if the existence of such sterile neutrinos is confirmed,
it would be necessary to reformulate the underlying models and find convincing physics reasons
for their existence and for the magnitude of their mixing. 

The well known LSND observation of the $\bar{\nu}_e$ appearance interpreted as the
$\bar{\nu}_{\mu} \rightarrow \bar{\nu}_e$ oscillation \cite{LSND} and the results of the
MiniBooNE experiment at similar $L/E_{\nu}$ \cite{MiniB} are consistent with oscillation
at a $\Delta m^2 \sim$ 1 eV$^2$. The so-called reactor anomaly, a deficit of the measured 
reactor $\bar{\nu}_e$ flux at $L \le 100$ m compared to the reevaluated flux prediction
\cite{Mueller,Mention}, as well as the analysis of the calibration of Gallex-SAGE solar neutrino 
detectors with the radioactive sources \cite{Giunti} have significance of $\sim 3\sigma$ and point
toward a similar interpretation. On the other hand, a number of short baseline oscillation 
searches have yielded null results, and a consistent and convincing picture of all of these 
effects has not been achieved as yet.

 Possible existence of $\sim$ 1 eV sterile neutrino on the effective neutrino Majorana
 mass $<m_{\beta\beta}>$ explored in the $0\nu\beta\beta$ decay would be profound. 
  $<m_{\beta\beta}>$ depends on the mass eigenstates
 neutrinos $\nu_i$ with masses $m_i$ as
 \begin{equation}
 <m_{\beta\beta}> = |\Sigma_i |U_{ei}|^2 m_i {\rm exp}(i\alpha_i) | ~,
 \label{eq:Majm}
  \end{equation}
  where $U_{ei}$ are the entries of the first row of the neutrino mixing matrix and $\alpha_i$
  are the so-called Majorana phases. 
  
  When the summation in Eq.~(\ref{eq:Majm}) goes only over the three known light $\nu_i$,
  as it is well known, in the case
  of the inverted hierarchy there is a lower limit   $<m_{\beta\beta}> \ge $ 20 meV.
  If the inverted hierarchy is shown to be the correct pattern of neutrino masses 
  experimentally (not by $0\nu\beta\beta$ decay), the above statement would imply
  that reaching the sensitivity to  $<m_{\beta\beta}> \sim $ 20 meV would represent
  a definitive test of the Majorana nature of neutrinos.
  The situation changes drastically when the existence of a fourth, necessarily sterile,
  neutrino is assumed. The lower limit of $<m_{\beta\beta}>$  disappears in that case; the
  whole region up to $<m_{\beta\beta}>$  $\sim$ 0.1 eV
  is now allowed. The best fit region for the normal hierarchy is changed substantially as well.
  
 \subsection{Where are we now}
 
 As pointed out in connection with Fig.~\ref{fig:history}, there has been a decade long hiatus
 in the improvements of the sensitivity of the searches for $0\nu\beta\beta$ decay when expressed
 as the upper limits of $<m_{\beta\beta}>$. The 2001 result of the Heidelberg-Moscow experiment,
 $1.9 \times 10^{25}$ years of the half-life for $^{76}$Ge, corresponds to $ <m_{\beta\beta}>$
 in the range 0.25-0.5 eV depending on the calculated value of the corresponding nuclear
 matrix element. Later, a subset of that collaboration reanalyzed the experimental result and
 claimed \cite{KK} a positive observation of the $0\nu\beta\beta$ decay with 
 $T_{1/2} = 2.23^{+0.44}_{-0.33} \times 10^{25}$ years, implying 
 $<m_{\beta\beta}>$ = 0.32 $\pm$ 0.03 eV using their preferred matrix elements ~\cite{Staudt}.
 That result remains unconfirmed.
 
\begin{figure}[tb]
 \begin{center}
\begin{minipage}[t]{8 cm}
\centerline{\epsfig{file=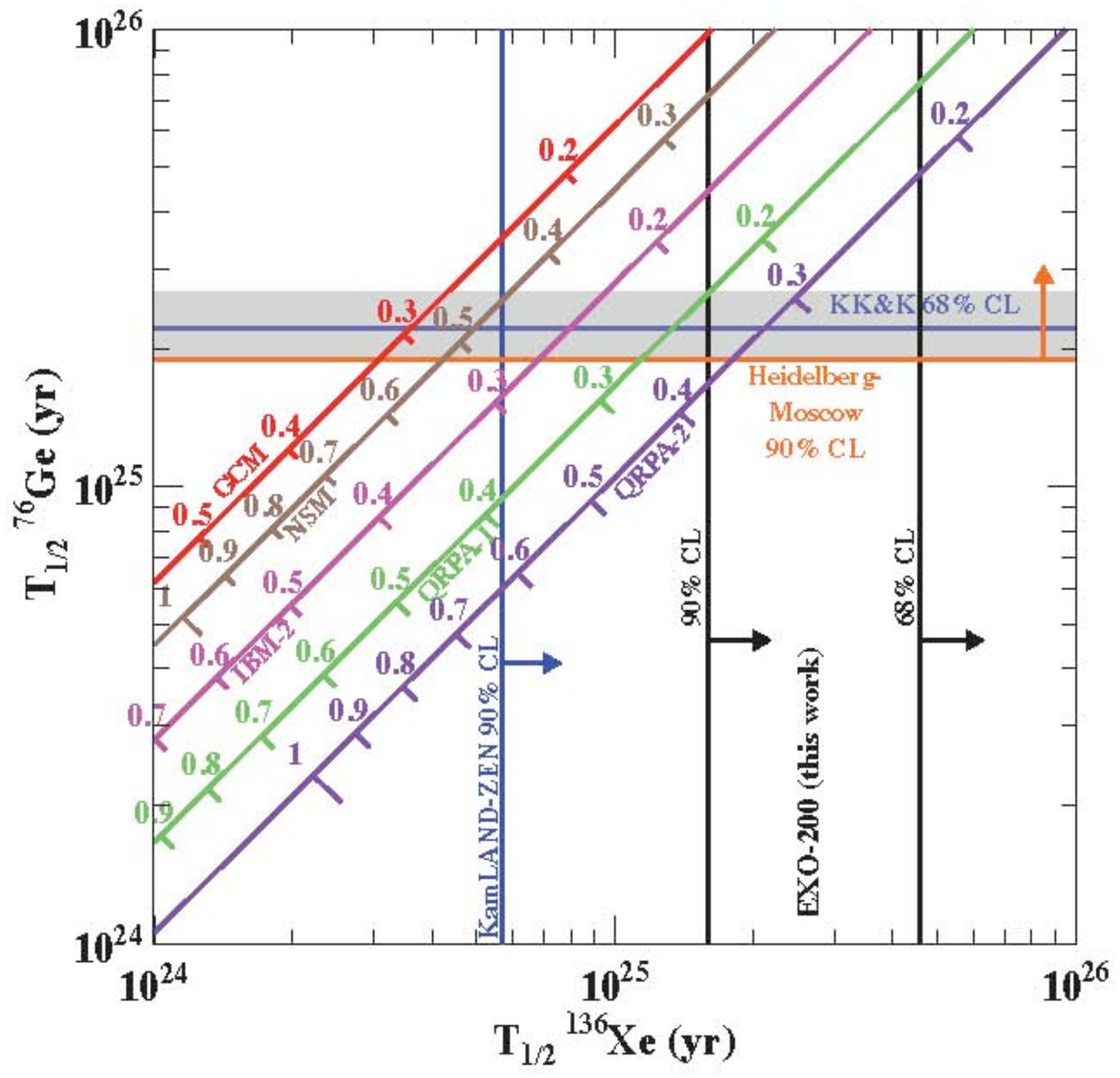,scale=0.6,angle=-90}}
\end{minipage}
\begin{minipage}[t]{16.5 cm}
\caption{Relation between the $0\nu\beta\beta$ half-lives of $^{136}$Xe and $^{76}$Ge.
The claim \cite{KK} is represented by the grey horizontal band along with the best limit
for $^{76}$Ge \cite{Heid-M}. The $^{136}$Xe limits from \cite{KamZ} and \cite{EXO200}
are shown by the vertical lines. Different theoretical calculations are represented
by the  diagonal lines, 
(GCM from \cite{Rod10}, NSM from \cite{Men09}, IBM-2 from \cite{Barea09},
RQRPA-1 from \cite{Sim09}, and QRPA-2 from \cite{Staudt})
with tick marks indicating the corresponding $<m_{\beta\beta}>$ values.}
\label{fig:me_plot}
\end{minipage}
\end{center}
\end{figure}

 Very recently two experimental searches for the $0\nu\beta\beta$ decay of $^{136}$Xe
 published their results. In Fig.~\ref{fig:me_plot} these recent results are compared with
 those for $^{76}$Ge. At 90\% CL the EXO-200 result corresponds to $<m_{\beta\beta}>$ 
  of less than 0.14 - 0.38 eV. At 68\% (90\%) CL it contradicts the claim \cite{KK}
  for all (most) matrix element element calculations (analogous figure appears in \cite{EXO200}) .

Both competing $^{136}$Xe experiments \cite{KamZ,EXO200} are continuing their runs and
substantial improvement of their half-life sensitivities both through accumulating
statistics and improving performance of their detectors is expected. Moreover,
other experiments of comparable size and sensitivity are expected to begin
their operation soon. The GERDA experiment \cite{Gerda} is using $^{76}$Ge
again, and will definitely confirm or reject the claim \cite{KK}. The CUORE
experiment \cite{Cuore} is using $^{130}$Te and will complement the
results obtained with $^{76}$Ge and $^{136}$Xe. There is, therefore, a realistic
chance that the upper edge of the so-called degenerate mass region 
($\langle m_{\beta\beta} \rangle$  $\sim$ 0.05 - 0.1 eV)
will be reached reasonably soon.

Even more ambitious proposals for ton (or multi-ton) size experiments
 to search for the
neutrinoless $\beta\beta$ decay exist. If the projected
performance (sensitivity, background suppression) could be 
achieved, and the corresponding funding secured, we should,
perhaps within a decade, reach the boundary of  $\langle m_{\beta\beta} \rangle$  $\sim$ 20  meV.
It is very important, essentially imperative, to continue the search for the 
$0\nu\beta\beta$ decay, by a variety of techniques
and using several candidate nuclei as sources. Only this way, if
the $0\nu\beta\beta$ decay is actually observed, we will have 
do doubts of its discovery. 

To extend the sensitivity beyond $\langle m_{\beta\beta} \rangle  \sim$   20  meV,
the region where only the normal hierarchy is present, appears to be
extremely challenging. No concrete plans exist for such experiments at the
present time.

Above, we used the effective Majorana mass    $<m_{\beta\beta}>$ to indicate the
sensitivity of the various searches for the $0\nu\beta\beta$ decay.
However, as pointed out already, this is not the only possible mechanism. Decay mediated by various
heavy particles is also a possibility and we have to keep it in mind when comparing the
$0\nu\beta\beta$ decay neutrino mass limits with the limits obtained in the direct 
kinematic experiments and from cosmology/astrophysics. 

As we tried to stress repeatedly in this article, the fundamental importance of the $0\nu\beta\beta$ decay
is the test of the total lepton number conservation. If discovered, it would be a crucial
step in the possible formulation of the `New Standard Model'.


\begin{thebibliography}{99}
\bibitem{Cuore} see crio.mib.infn.it/wigmi/pages/cuore.php
\end{thebibliography}
\end{document}